\documentclass{aa}

\usepackage{graphicx}
\usepackage{txfonts}
\usepackage{amsmath}
\usepackage{xcolor}
\newcommand{\etamaxpct}{2.5}
\newcommand{\rhowind}{0.020^{+0.018}_{-0.011}}
\newcommand{\rhostarburst}{0.26^{+0.37}_{-0.16}}
\newcommand{\rhoratio}{13.5}
\newcommand{\rhoratiolo}{5.3}
\newcommand{\rhoratiohi}{37}
\newcommand{\Pgtone}{0.998}
\newcommand{\Pgtthree}{0.95}
\newcommand{\nwind}{7}
\newcommand{\nstarburst}{50}
\newcommand{\Pcontrast}{0.93}
\newcommand{\etawind}{5.2\times10^{-4}}
\newcommand{\twostackratio}{1.75}
\newcommand{\twostacklo}{0.26}
\newcommand{\twostackhi}{24}
\newcommand{\sigmarho}{0.3}

\newcommand{\sfxray}{16}
\newcommand{\contrastindep}{13.4}
\newcommand{\contrastindeplo}{6.1}
\newcommand{\contrastindephi}{30}
\newcommand{\Pcontrastthree}{0.98}
\newcommand{\xiratioone}{0.07}
\newcommand{\xiratiothree}{0.22}
\newcommand{\Neff}{3.4}
\newcommand{\topweight}{47}
\newcommand{\sigmaneg}{0.7}
\newcommand{\sigmapos}{0.01}
\newcommand{\sigmasoft}{0.5}
\newcommand{\compatlow}{4}
\newcommand{\compathigh}{55}
\newcommand{\rhoresampled}{0.003}
\newcommand{\rhoshiftdex}{0.8}
\newcommand{\nattenpc}{72}
\newcommand{\Rerase}{14}
\newcommand{\jensen}{11}

\newcommand{\rhowindmin}{0.24}
\newcommand{\contrastmin}{1.1}
\newcommand{\contrastmax}{927}
\newcommand{\nlowerlim}{4}
\newcommand{\rholinuniform}{6\times10^{-4}}
\newcommand{\pctcontrastgtthree}{99.9}
\newcommand{\Fteva}{1.8\times10^{-13}}
\newcommand{\Ftevb}{9.6\times10^{-14}}
\newcommand{\eblsupp}{8}
\newcommand{\tauonetev}{0.05}
\newcommand{\tautentev}{0.19}
\newcommand{\Lpal}{3.2\times10^{39}}
\newcommand{\ratioUFOPAL}{32}
\newcommand{\ratioFGRPAL}{10}
\newcommand{\spandex}{1.5}
\newcommand{\rhowindnoNfourone}{0.017}
\newcommand{\TSnoNfourone}{15.1}
\usepackage[colorlinks=true, citecolor=blue, linkcolor=blue, urlcolor=blue]{hyperref}

\begin{document}

\title{Gamma-ray calorimetry of Seyfert winds}

\titlerunning{Gamma-ray calorimetry of Seyfert winds}
\authorrunning{Domínguez}

\author{Alberto Dom\'inguez\inst{1,2}\fnmsep\thanks{alberto.d@ucm.es}}

\institute{Department of EMFTEL, Universidad Complutense de Madrid, E-28040 Madrid, Spain
           \and
           Instituto de Física de Partículas y del Cosmos (IPARCOS), Universidad Complutense de Madrid, E-28040 Madrid, Spain}

\date{Received ...; accepted ...}

\abstract
{Two stacking analyses with the \textit{Fermi} Large Area Telescope (LAT) detected GeV emission from local radio-quiet Seyferts, one selected by an X-ray ultra-fast outflow (UFO) and one hard-X-ray-bright, with similar luminosities attributed to different mechanisms: wind-driven shocks and the accretion-disk corona.}
{We investigate how efficiently Seyfert outflows convert mechanical power into GeV photons, whether the two detections require distinct mechanisms, and how Seyferts compare with low-ionization nuclear emission-line regions (LINERs).}
{We perform no new reduction of \textit{Fermi}-LAT data, but homogenize published measurements to 0.1--100~GeV, place local mechanical accelerators on one $L_\gamma$--$\dot{E}$ diagram, and express $\eta=L_\gamma/\dot{E}$ as a fraction $\rho$ of the fiducial complete-calorimetry efficiency $\eta_{\max}\simeq\etamaxpct\%$, inverting $\rho$ for the ambient density required by $pp$ losses.}
{Winds from Seyfert active galactic nuclei (AGN) have $\rho=\rhowind$ for an acceleration efficiency of $0.1$, a factor of $\simeq\rhoratio$ below starburst superwinds ($\rho=\rhostarburst$). The difference exceeds a factor of three with probability $\Pgtthree$ and is independent of acceleration parameters common to both classes; it is conditional on the wind kinetic powers, collapsing only if every wind sits at its minimum launch radius. The implied densities, $\sim10$ and $\sim10^2$~cm$^{-3}$, are consistency values, as density, shock radius and speed are degenerate. The two stacking detections agree within a factor of $\twostackratio$ in $\gamma$-ray output per unit Eddington luminosity, but the 95 percent interval, $\twostacklo$--$\twostackhi$, is too wide to distinguish them. A third stack, of 186 low-luminosity AGN, extends the comparison 1.5~dex lower, ordered by AGN power.}
{GeV detectability of a non-jetted AGN is set by the product of mechanical power and calorimetric fraction. LINER outflows fail on power; for the adopted wind energetics, UFO-selected Seyfert winds fail on calorimetry; starburst superwinds succeed on both.}


\maketitle
\nolinenumbers

\section{Introduction}
\label{sec:intro}

Seyfert galaxies are the archetypal radio-quiet active galactic nuclei (AGN) of the local Universe. Lacking the powerful relativistic jets that make blazars and radio galaxies the dominant extragalactic $\gamma$-ray population \citep{Ajello2020c}, they were long expected to be $\gamma$-ray dark, and three decades of searches with the Energetic Gamma Ray Experiment Telescope (EGRET) and the \textit{Fermi} Large Area Telescope (LAT) returned only non-detections \citep{Lin1993, Cillis2004, Teng2011, Ackermann2012a}. That situation has changed over the last five years, with three stacking measurements pointing to different physical interpretations.

First, \citet{Ajello2021} stacked 11.1~yr of LAT data at the positions of eleven nearby radio-quiet Seyferts selected because their X-ray spectra show an ultra-fast outflow \citep[UFO; $v>0.1c$;][]{Tombesi2010a, Gofford2013}, and detected the sample at $5.1\sigma$. They interpreted the signal as $\pi^0$-decay emission from cosmic rays (CRs) accelerated at the forward shock driven by the wind into the interstellar medium (ISM), and reported that the outflows transfer $\approx4\times10^{-4}$ of their mechanical power to $\gamma$ rays. Second, \citet{Liu2025} stacked 15~yr of LAT data for 37 nearby, hard-X-ray-bright, radio-quiet Seyferts drawn from the \textit{Swift} Burst Alert Telescope AGN Spectroscopic Survey \citep[BASS;][]{Koss2022}, detected them at $5.2\sigma$, and interpreted the emission as inverse-Compton radiation from non-thermal electrons in the accretion-disk corona --- requiring, in order to escape $\gamma\gamma$ absorption, an ``extended corona'' of radius $\sim2.7\times10^6\,R_{\rm g}$, i.e. a few parsecs.

The third probes the opposite end of the AGN luminosity function. \citet{Karwin2026} stacked 14.4~yr of LAT data for 186 subthreshold low-luminosity AGN (LLAGN) from the Palomar spectroscopic survey \citep{Ho2008}, detecting the full sample at $5.2\sigma$ and its 40-source Seyfert subset at $2.9\sigma$; earlier work on the same parent sample had yielded only four individual detections \citep{deMenezes2020}. They find the stacked signal to follow the relation between $\gamma$-ray and infrared (IR) luminosity, $L_\gamma$--$L_{\rm IR}$, of star-forming galaxies, to come almost entirely from spiral hosts, and to vanish in an emission-line-free control sample.

The first two have comparable statistical significance and average luminosity, $\sim10^{40}$~erg~s$^{-1}$, but they attribute the emission to emitting regions that differ in size by more than an order of magnitude and to entirely different radiative processes. Neither analysis was designed to test the other's hypothesis. Meanwhile, a third possibility --- CRs accelerated by star formation in the host --- is the standard explanation for the nearest individually GeV-detected radio-quiet Seyferts, NGC~1068, NGC~4945 and the Circinus galaxy \citep{Lenain2010, Ackermann2012b, Ajello2020, Guo2019}, all of which are composite starburst--AGN systems, and NGC~1068 is additionally a candidate high-energy neutrino source \citep{Ajello2023}. A fourth radio-quiet Seyfert, the Compton-thick NGC~3281 at $48$~Mpc, has since been reported at $6.22\sigma$ with a luminosity of $(5.9\pm1.7)\times10^{41}$~erg~s$^{-1}$ and a soft spectrum, $\Gamma=2.61\pm0.24$, which star formation cannot supply \citep{LiuJR2025}; we return to it in Sect.~\ref{sec:sfceiling}.

We recently constrained the complementary question at the low-luminosity end of the AGN population. In \citet[][hereafter D26]{Dominguez2026} we combined integral-field kinematics, obtained at the Gran Telescopio Canarias (GTC) with the Multi-Espectr\'ografo en GTC de Alta Resoluci\'on para Astronom\'{\i}a (MEGARA), of six local low-ionization nuclear emission-line regions (LINERs) with 17~yr of LAT data and showed that their spatially resolved ionized outflows cannot power the observed high-energy emission of the most efficient $\gamma$-ray LINERs: the outflows are too weak by three to four orders of magnitude, which requires a compact nuclear jet. Seyferts differ from LINERs in almost every relevant respect --- radiatively efficient, not advection-dominated, accretion, one to three decades more bolometric power, and nuclear winds that are faster by two orders of magnitude. The same question therefore has a different answer for Seyferts, and addressing it requires a framework in which the two populations can be compared on the same physical footing.

That framework is calorimetry. For any mechanical outflow that accelerates protons at a shock, the $\gamma$-ray luminosity cannot exceed
\begin{equation}
L_\gamma^{\max}=\xi_{\rm CR}\,f_{\pi^0}\,f_{\rm band}\,\dot{E}\equiv\eta_{\max}\dot{E},
\label{eq:ceiling}
\end{equation}
where $\dot{E}$ denotes the mechanical power of whichever outflow is under discussion, $\dot{E}_{\rm K}$ being reserved for the kinetic power of a UFO, $\xi_{\rm CR}$ is the fraction of that power channelled into relativistic protons, $f_{\pi^0}$ the fraction of the energy lost in inelastic $pp$ collisions that emerges as $\pi^0$-decay photons, and $f_{\rm band}$ the fraction of that emission falling in the observed band. Throughout, ``max'' denotes complete $pp$ calorimetry for the adopted acceleration normalization, not an absolute physical maximum, since $\xi_{\rm CR}$ is not bounded at its fiducial value (Sect.~\ref{sec:ceiling}). The observed efficiency $\eta=L_\gamma/\dot{E}$ therefore defines a physically meaningful quantity, the calorimetric ratio $\rho=\eta/\eta_{\max}$, which is the fraction of the injected CR energy actually radiated before the protons escape. This is the same quantity that regulates the $L_\gamma$--star-formation-rate relation of star-forming galaxies \citep{Kornecki2020}, and it can be computed from first principles once the density, size and speed of the shocked region are known.

In this paper we apply that framework to the local AGN population. Section~\ref{sec:samples} defines the samples; Sect.~\ref{sec:framework} sets out the homogenization and the calorimetric formalism; Sect.~\ref{sec:results} presents the efficiencies and the densities they imply; Sect.~\ref{sec:two} asks whether the two stacking detections are the same measurement; Sect.~\ref{sec:discussion} discusses the physical picture, the comparison with LINERs, and what would falsify it. Throughout we adopt $H_0=70$~km~s$^{-1}$~Mpc$^{-1}$, $\Omega_{\rm m}=0.3$, $\Omega_\Lambda=0.7$.

Every input used here is a published measurement; the contribution of this work is to homogenize these measurements, to construct the calorimetric analysis, and to perform the tests that this framework enables.

\section{Samples}

\begin{table*}
\caption{The S-WIND sample: local radio-quiet Seyfert galaxies hosting an X-ray ultra-fast
outflow, ordered by predicted detectability.}
\label{tab:swind}
\centering
\small
\setlength{\tabcolsep}{4pt}
\begin{tabular}{lccccccccc}
\hline\hline
Source & Type & $z$ & $D_{\rm L}$ & $\log\dot{E}_{\rm K}$ & $\log L_{\rm bol}$ &
$F_\gamma^{95\%}$ & $\log L_\gamma^{\rm pred}$ & $\rho^{95\%}$ & $F^{\rm pred}/F^{95\%}$\\
 & & & (Mpc) & (erg s$^{-1}$) & (erg s$^{-1}$) & & (erg s$^{-1}$) & & \\
\hline
NGC 7582 & Sy2 & 0.0052 & 22.4 & 44.15 & 43.30 & 4.7 & 40.86 & 0.01 & 2.34 \\
NGC 5506 & Sy1.9 & 0.0060 & 25.8 & 44.00 & 44.30 & 6.4 & 40.71 & 0.02 & 0.91 \\
MCG-5-23-16 & Sy2 & 0.0084 & 36.2 & 43.50 & 44.10 & 4.3 & 40.21 & 0.09 & 0.22 \\
SWIFT J2127.4+5654 & Sy1 & 0.0140 & 60.6 & 44.20 & 44.50 & 9.1 & 40.91 & 0.11 & 0.18 \\
PG 1211+143 & Sy1 & 0.0810 & 368.1 & 45.30 & 45.10 & 3.7 & 42.01 & 0.13 & 0.15 \\
Mrk 290 & Sy1 & 0.0300 & 131.4 & 44.35 & 44.40 & 4.5 & 41.06 & 0.18 & 0.11 \\
Ark 120 & Sy1 & 0.0330 & 144.9 & 44.65 & 44.60 & 7.5 & 41.36 & 0.18 & 0.11 \\
MR 2251-178 & Sy1 & 0.0640 & 287.4 & 45.00 & 45.80 & 7.4 & 41.71 & 0.32 & 0.06 \\
NGC 4151 & Sy1 & 0.0033 & 14.2 & 42.50 & 43.40 & 10.6 & 39.21 & 0.35 & 0.06 \\
NGC 4507 & Sy2 & 0.0120 & 51.9 & 42.90 & 44.30 & 3.4 & 39.61 & 0.60 & 0.03 \\
Mrk 509 & Sy1 & 0.0340 & 149.4 & 44.20 & 44.70 & 9.5 & 40.91 & 0.70 & 0.03 \\
\hline
\end{tabular}
\tablefoot{Columns 2--7 are taken directly from \citet{Ajello2021}: Seyfert type, redshift,
kinetic power (geometric mean of their tabulated minimum and maximum), bolometric luminosity,
and the 95\% confidence photon-flux upper limit over 1--800~GeV in units of
$10^{-11}$~ph~cm$^{-2}$~s$^{-1}$ ($\Gamma=2.0$). $D_{\rm L}$ is computed from $z$ for
$H_0=70$~km~s$^{-1}$~Mpc$^{-1}$, $\Omega_{\rm m}=0.3$. Column 8 is the wind-powered
luminosity predicted by applying the population-average efficiency
$\eta_{\rm wind}=5.2\times10^{-4}$ to each source; column 9 is the calorimetric
ratio allowed by that source's own upper limit; column 10 is the ratio of the predicted to the
limiting energy flux, both in 0.1--100~GeV.}
\end{table*}

\label{sec:samples}

We use five populations, each defined by a published, quantitative and reproducible selection. We did not assemble any of the samples ourselves; this is deliberate, because it ensures that the selection functions are those already established in the literature and that our comparison is not tuned to the present analysis.

\paragraph{S-WIND.} The eleven radio-quiet Seyferts of \citet{Ajello2021}: sources with a UFO detected in blueshifted Fe~K absorption \citep{Reeves2003, Tombesi2010a, Tombesi2012, Gofford2013}, restricted to $z<0.1$ and $v>0.1c$, and radio-quiet according to the original references. Their properties are reproduced in Table~\ref{tab:swind}. The sample spans $\log L_{\rm bol}$(erg s$^{-1}$)$=43.3$--$45.8$ with a mean of $44.4$, black-hole masses $\log(M_\bullet/M_\odot)=6.4$--$8.7$, and a median redshift of $0.014$. Kinetic powers are the geometric mean of the minimum and maximum values tabulated by \citet{Ajello2021}, which correspond to the minimum and maximum plausible launch radii; the half-width of that interval, $0.6$--$1.7$~dex, dominates the error budget on $\dot{E}_{\rm K}$ and is propagated as such. The sample value quoted by \citet{Ajello2021}, $1.8\times10^{44}$~erg~s$^{-1}$, is the mean kinetic power against which their stacked luminosity is divided to obtain the quoted efficiency; it is thus a ratio of sample means, and Appendix~\ref{app:weights} sets out what that estimator constrains.

\paragraph{S-XRAY.} The 37 sources of the ``faint $\gamma$-ray'' sample of \citet{Liu2025}: AGN from the second BASS data release that are non-blazar, radio-quiet by both the hard-X-ray radio-loudness parameter $\mathcal{R}_{\rm rX}<10^{-4}$ and the classical $\mathcal{R}_{\rm rB}<10$, not classified as star-forming on the Baldwin--Phillips--Terlevich diagram, within 60~Mpc, and with $F_{14-195\,\rm keV}>2\times10^{-11}$~erg~cm$^{-2}$~s$^{-1}$. The sample is listed in Table~3 of \citet{Liu2025}, which we do not reproduce here. None is individually detected: the test statistic (TS) is below 25 for all 37.

\paragraph{S-DET.} The three radio-quiet Seyferts individually detected in GeV $\gamma$ rays that appear in the homogeneous compilation of \citet{Kornecki2020}: NGC~1068, NGC~4945 and Circinus. We take their 0.1--100~GeV luminosities, infrared luminosities, star-formation rates and distances from that work, which recomputes all quantities on a common, self-consistent distance scale --- necessary for the calorimetric test in Sect.~\ref{sec:sfceiling}.

\paragraph{L-OUT and comparison objects.} The six LINERs with MEGARA/GTC-resolved ionized-outflow energetics and the four literature comparison objects (M82, NGC~253, M87, Cen~A) of D26.

\paragraph{S-PAL.} The 40 Seyfert nuclei in the subthreshold Palomar LLAGN sample of \citet{Karwin2026}, stacked over 1--800~GeV. These are optically selected and of much lower AGN luminosity than either sample above, and none is individually detected. Black-hole masses and distances are not tabulated for the subset; S-PAL thus enters only at the population level. The same work also stacks 84 Palomar LINERs.

We also use the two published control samples: the 20 nearby radio-quiet AGN in which a UFO was searched for and not found \citep{Ajello2021}, and the 27 X-ray-faint BASS Seyferts of \citet{Liu2025}. Neither is detected in $\gamma$ rays.

The two samples are not, however, strictly independent. \citet{Liu2025} state that their sample does not overlap with that of \citet{Ajello2021}, and use this as one argument that a wind origin for their signal can be excluded. Cross-matching the two published tables by position, we find that two objects are in fact common to both: MCG--5-23-16 ($\alpha,\delta=146.92,-30.95$) and NGC~4507 ($188.90,-39.91$). The overlap is small --- 2 of 11 and 2 of 37 --- and does not by itself undermine either detection: removing both sources changes the mean $\log M_\bullet$ of S-WIND by $+0.13$~dex and of S-XRAY by $-0.02$~dex, well within the uncertainties of everything derived below. A third source, NGC~7172, is common to the two studies in a different way: it is one of the twenty objects that \citet{Ajello2021} searched for a UFO and used as a control precisely because none was found, and it is a primary target of \citet{Liu2025}, where it contributes $TS=3.7$. This case provides the opposite situation, since here a dedicated search was made and returned nothing: at least one member of S-XRAY is known not to host a detectable UFO. Taken together, the three cases mean that the two measurements are not strictly independent, and that the absence of overlap therefore cannot be used as evidence against a common origin.

One source in S-WIND warrants a caveat. \citet{Karwin2026} exclude NGC~4151 from their sample because its $\gamma$-ray emission is confused with that of the nearby blazar 1E~1217.9+3945. \citet{Peretti2025} show that the localisation of 4FGL~J1210.3+3928 \citep{Ballet2023} may be affected by the superposition of the two sources, while arguing that particle acceleration at a UFO wind termination shock can account for the observed luminosity and spectral shape. NGC~4151 also dominates the S-WIND stacked significance. We quantify the consequences in Sect.~\ref{sec:third}; they are modest, but the ambiguity remains relevant to the interpretation of S-WIND.

More importantly, the absence of a detected UFO does not imply the absence of a wind. UFOs are common in nearby AGN observed with sufficient X-ray sensitivity \citep[e.g.,][]{Tombesi2010b, Gofford2013}, their detection depends on the line of sight intercepting the outflow and on the epoch observed, and the S-XRAY selection never tested for one. Whether the two samples can nonetheless be treated as differently weighted draws from a common population of nearby, hard-X-ray-bright, radio-quiet Seyferts is a question we test, not one we assume: the answer, given in Sect.~\ref{sec:two} and Appendix~\ref{app:robust}, is that they are matched in black-hole mass but not in distance or AGN luminosity, and every comparison between them in this paper is qualified accordingly.

\section{Method}
\label{sec:framework}

\subsection{Homogenization}
\label{sec:homog}

The published measurements are reported in four different energy bands (1--800~GeV, 1--300~GeV, 0.05--500~GeV, 0.1--100~GeV). We convert all of them to 0.1--100~GeV, the energy band adopted by the star-forming-galaxy compilation that provides our star-formation calorimeter, by integrating the published power law,
\begin{equation}
k=\frac{\int_{0.1}^{100}E^{1-\Gamma}\,{\rm d}E}{\int_{E_1}^{E_2}E^{1-\Gamma}\,{\rm d}E},
\end{equation}
using each measurement's own photon index. The correction factors are $k\simeq1.3$ for S-WIND ($\Gamma=2.1$), $k\simeq2.2$ for S-XRAY ($\Gamma=2.32$), $k\simeq0.77$ for the LINER limits ($\Gamma=2.2$) and $k\simeq2.0$ for S-PAL ($\Gamma=2.3$); we quote two significant figures because $k$ can vary by up to 20 percent within the published uncertainty on $\Gamma$. Extrapolation beyond the measured band is a model assumption; its effect is quantified in Appendix~\ref{app:mc} by varying $\Gamma$ over its published uncertainty, which changes $k$ by at most 20 percent and does not affect any conclusion.

Attenuation by the extragalactic background light (EBL) is negligible over the reference band. The samples are local: S-WIND spans $z=0.0033$--$0.081$ with a median of $0.014$, S-XRAY lies within $60$~Mpc, and the individually detected Seyferts are within $11$~Mpc. At these distances the optical depth is small throughout the reference band, and the homogenized luminosities of Sect.~\ref{sec:rho} are therefore unaffected, as in D26 \citep[e.g.,][]{SaldanaLopez2021}. Two points warrant qualification. The first is the upper end of the $1$--$800$~GeV band over which the S-WIND and S-PAL fluxes are quoted, where attenuation is no longer strictly zero for the most distant source in the sample, PG~1211+143 at $z=0.081$; because our band correction integrates a power law over $0.1$--$100$~GeV, the effect on the reference-band luminosity is smaller still. The second is the TeV extrapolation of Sect.~\ref{sec:falsify}, where attenuation is no longer negligible; there we apply the optical depths of the \citet{SaldanaLopez2021} model directly, and at these distances they suppress the $1$--$10$~TeV band flux by $\eblsupp$ percent ($\tau=\tauonetev$ at $1$~TeV and $\tautentev$ at $10$~TeV for NGC~7582). Neither affects the calorimetric results.

Two independent checks provide validation of the procedure. Applying it to the S-XRAY measurement over 0.1--800~GeV returns $3.5\times10^{40}$~erg~s$^{-1}$, against the $3.6\times10^{40}$~erg~s$^{-1}$ quoted by \citet{Liu2025} for the same conversion. Applying it to the M82 luminosity of D26 returns a value within 15 percent of the independently derived \citet{Kornecki2020} luminosity for the same galaxy, the residual difference being due to the different adopted distances and fluxes.

\subsection{The calorimetric ceiling}
\label{sec:ceiling}

We adopt $\xi_{\rm CR}=0.1$, the canonical diffusive-shock-acceleration value used both in supernova-remnant modeling and in the CR-transport models of star-forming galaxies \citep[e.g.,][]{Caprioli2014, Kornecki2020}; $f_{\pi^0}=1/3$, the fraction of the proton energy lost to $\pi^0$ decay in inelastic $pp$ collisions \citep{Kelner2006}; and $f_{\rm band}=0.76$, the fraction of the resulting emission that falls in 0.1--100~GeV for a proton spectrum of index $\simeq2.2$ \citep{Kornecki2020}. This gives
\begin{equation}
\eta_{\max}=2.5^{+1.5}_{-0.9}\times10^{-2},
\end{equation}
where the uncertainty is the 68 percent Monte Carlo interval obtained by varying $\xi_{\rm CR}$ log-uniformly over $0.05$--$0.20$, $f_{\pi^0}$ uniformly over $0.30$--$0.37$ and $f_{\rm band}$ uniformly over $0.65$--$0.85$ (Appendix~\ref{app:mc}). Applying Eq.~(\ref{eq:ceiling}) to supernova mechanical power reproduces the independent star-formation calorimetric limit of \citet{Kornecki2020} to within 14 percent, providing an external check on the normalization.

Two aspects of Eq.~(\ref{eq:ceiling}) warrant clarification. First, $\xi_{\rm CR}=0.1$ is a typical value inferred for supernova-remnant shocks, not an upper bound: values approaching $0.3$ are discussed in the diffusive-shock-acceleration literature \citep{Caprioli2014}, and $\eta_{\max}$ would then be $\simeq7.5\times10^{-2}$. Every absolute value of $\rho$ quoted below therefore scales as $\xi_{\rm CR}^{-1}$ and should be interpreted as ``$\rho$ for $\xi_{\rm CR}=0.1$''. Second, and this is why we work with $\rho$ at all, the ratio of $\rho$ between two classes is exactly independent of $\xi_{\rm CR}$, $f_{\pi^0}$ and $f_{\rm band}$, because they cancel; this is verified numerically (Appendix~\ref{app:robust}). The cancellation is algebraic, but its physical interpretation requires the same values to apply to both classes; an AGN wind shock and a supernova shock are different environments, and this condition is not automatic. Appendix~\ref{app:xidiff} relaxes the assumption: drawing the three parameters independently for the two classes leaves the median contrast unchanged with $P(\rho_{\rm starburst}>3\,\rho_{\rm wind})=\Pcontrastthree$, and the contrast falls below three only if AGN-wind shocks are a factor $\simeq4.5$ less efficient accelerators than supernova shocks. The comparisons are therefore robust in a way that the absolute values are not, but the robustness is conditional and we phrase the conclusions accordingly.

\subsection{Calorimetric fraction and ambient density}
\label{sec:fcal}

Protons injected at a shock lose energy to inelastic $pp$ collisions on a timescale $t_{pp}=(n\,c\,\sigma_{pp}\kappa)^{-1}$, with $\sigma_{pp}\simeq34$~mb and inelasticity $\kappa=0.5$ \citep{Kelner2006}, i.e. $t_{pp}\simeq6.2\times10^{7}\,(n/{\rm cm^{-3}})^{-1}$~yr. They remain confined to the shocked shell for a time of order the shock age, $t_{\rm res}\simeq R/v_{\rm sh}$. The fraction of the injected CR energy that is radiated is then
\begin{equation}
f_{\rm cal}=1-\exp\!\left(-\frac{t_{\rm res}}{t_{pp}}\right),
\label{eq:fcal}
\end{equation}
from which, provided the emission is purely hadronic and $\xi_{\rm CR}$ takes its fiducial value,
\begin{equation}
\rho=f_{\rm cal}.
\label{eq:rhofcal}
\end{equation}
Equation~(\ref{eq:rhofcal}) is the only step in this paper that directly identifies an observed ratio with a physical fraction, and it holds under two conditions. If $\xi_{\rm CR}>0.1$, then $\rho<f_{\rm cal}$ and the densities inferred below are underestimates. If $p\gamma$ interactions contribute to the observed emission \citep[e.g.,][]{Inoue2019, Murase2024}, then part of $\rho$ is not $pp$ and the inferred densities become upper limits. In both cases the inversion should therefore be regarded as a consistency test --- ``a medium of this density suffices'' --- and not as a measurement, and only the calorimetric ratios themselves, which involve no such identification, are used in the comparisons of Sect.~\ref{sec:rho}. This decomposition of the $\gamma$-ray output of an AGN-driven shock into an acceleration efficiency and a proton calorimetric fraction follows \citet{Lamastra2016}, who applied it to the circumnuclear outflow of NGC~1068 and found that the wind can account for the whole of its GeV emission only for $\eta_p\gtrsim0.3$ and $F_{\rm cal}\gtrsim0.5$. Here we use the decomposition in the opposite direction: instead of fitting a single object, we take the measured population efficiency and solve for the calorimetric fraction it implies. Equation~(\ref{eq:fcal}) can then be inverted to give the ambient density required to reproduce a given observed $\rho$. For the AGN winds we adopt $R=100$~pc and $v_{\rm sh}=500$~km~s$^{-1}$, the mid-range of the forward-shock radius ($20$--$300$~pc) and velocity ($200$--$1000$~km~s$^{-1}$) inferred for UFO-driven bubbles \citep[e.g.,][]{King2010, Zubovas2012, Liu2018, Ajello2021}; for starburst superwinds we adopt $R=100$~pc and $v_{\rm sh}=600$~km~s$^{-1}$, appropriate to the nuclear CR cooling region \citep[e.g.,][]{Strickland2009, Kornecki2020}. Both ranges are varied in the Monte Carlo. We stress that $R$ and $v_{\rm sh}$ are model-dependent quantities adopted from the literature and not directly measured in this work, and that the inferred densities inherit that dependence; the calorimetric ratios themselves do not.

\section{Results}
\label{sec:results}

\subsection{The mechanical-feedback \texorpdfstring{$\gamma$}{gamma}-ray diagram}
\label{sec:diagram}

\begin{figure}[t]
\centering
\includegraphics[width=\columnwidth]{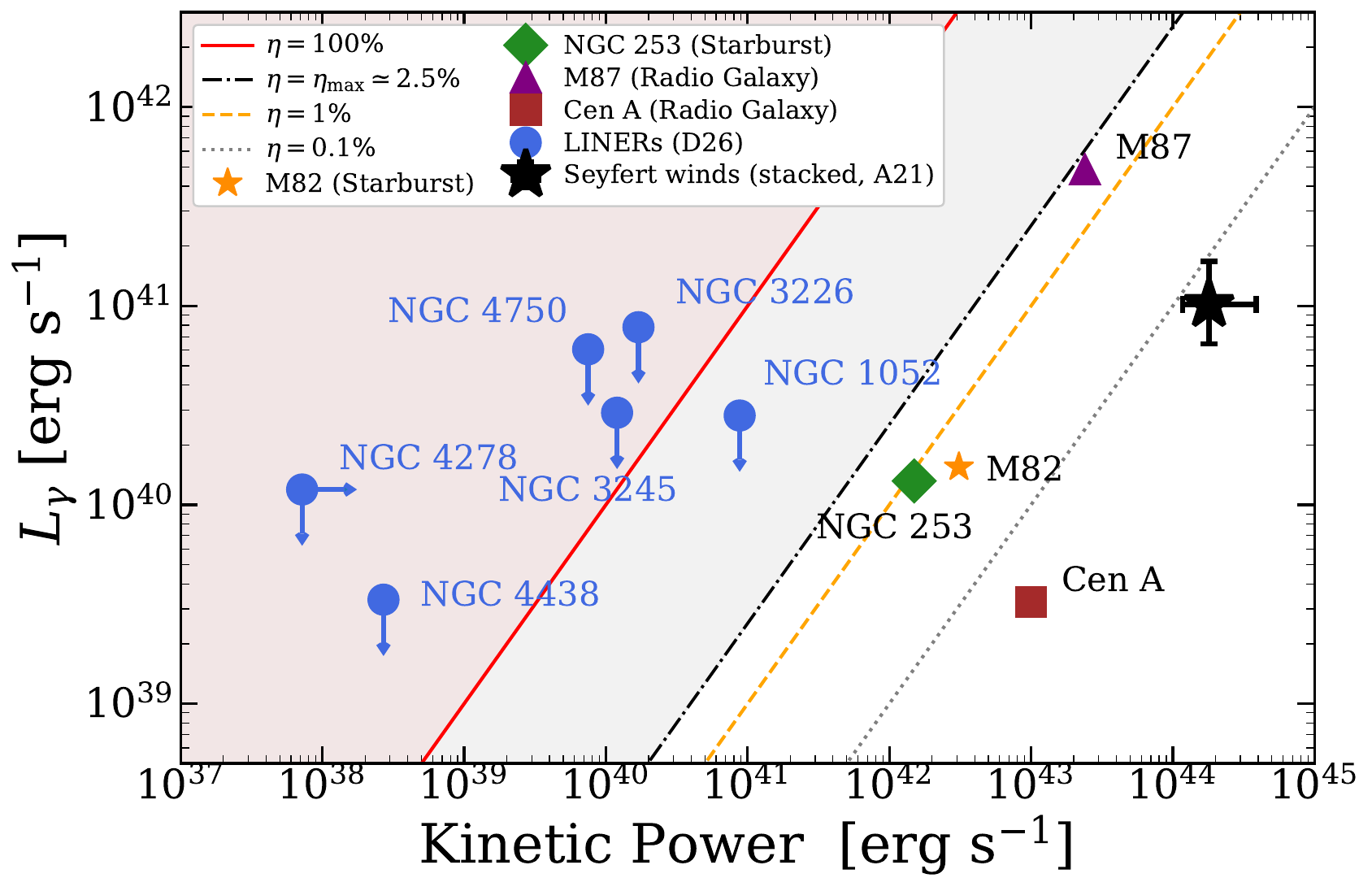}
\caption{Gamma-ray luminosity against mechanical power for local accelerators. Diagonal lines are contours of constant apparent efficiency $\eta=L_\gamma/\dot{E}$. The dark shading is the strictly unphysical region $\eta>1$ used in Fig.~1 of D26; the light shading is the region above the hadronic calorimeter ceiling $\eta_{\max}\simeq\etamaxpct\%$ adopted here. All six D26 LINERs are labeled, as in that paper: open circles with downward arrows are their 95\% confidence limits, and the rightward arrow on NGC~4278 marks a kinetic power that is itself a lower limit. NGC~1052 is the only LINER below $\eta=1$, and the only one yielding a physically meaningful constraint. The black hexagon, deliberately distinct from the markers used for individual objects, is the stacked Seyfert AGN wind measurement of \citet[][A21]{Ajello2021}; its horizontal bar spans the logarithmic to the linear sample mean of the published minimum and maximum kinetic powers. It is a population measurement and is not commensurate with the individual objects around it: it carries no distance and represents eleven sources through a single fitted efficiency. All luminosities are homogenized to 0.1--100~GeV.}
\label{fig:master}
\end{figure}

Figure~\ref{fig:master} places every local mechanical accelerator with both a measured $\gamma$-ray luminosity (or limit) and a measured mechanical power on a single plane, extending the diagnostic diagram of D26 by almost an order of magnitude in $\dot{E}$ and, more importantly, by an entire AGN class. Diagrams of this kind --- a $\gamma$-ray luminosity against an energy-injection proxy, with diagonal contours of constant efficiency and a calorimetric ceiling --- are long established for star-forming galaxies, where the abscissa is the infrared luminosity or the star formation rate \citep[e.g.,][]{Ackermann2012b, Kornecki2020, Karwin2026}; Figure~\ref{fig:master} extends this approach by placing AGN winds, superwinds, LINER outflows, and jets on a common plane in terms of mechanical power. Two boundaries are drawn. The dark shading marks $\eta>1$, the strictly unphysical region of D26; the light shading marks the region above the hadronic calorimeter ceiling, $\eta>\eta_{\max}$. We show both because the ceiling adopted here is a factor of $40$ below the boundary used by D26; consequently, a source that lies comfortably below the unphysical line there --- NGC~1052 is the example --- lies above the ceiling here. A measured luminosity cannot exceed the ceiling for $\xi_{\rm CR}=0.1$; an upper limit may fall there, and simply means that the limit is not yet constraining.

Three features are immediately apparent. All six LINER upper limits lie above the ceiling, which is another way of stating what D26 concluded: their ionized outflows are too weak by several orders of magnitude for the LAT limits to be constraining. Only NGC~1052 --- the radio-loud control in that paper and the only one of the six below $\eta=1$ --- yields a physically meaningful constraint ($\rho<13$). The starburst superwinds and the Seyfert winds both lie below the ceiling, but by very different margins. And the radio galaxies M87 and Cen~A bracket the whole range, because their GeV emission is not of hadronic calorimetric origin; we include them only as reference points and do not interpret their $\rho$.

\subsection{Seyfert winds are poor calorimeters}
\label{sec:rho}

The stacked efficiency of \citet{Ajello2021}, $\eta=4.0^{+2.3}_{-2.0}\times10^{-4}$ over 1--800~GeV, becomes $\eta=\etawind$ in our reference band, which gives
\begin{equation}
\rho_{\rm wind}=\rhowind,
\end{equation}
where the interval includes both the measurement uncertainty and the uncertainty on $\eta_{\max}$. Seyfert AGN winds radiate about 2\% of the $\gamma$-ray luminosity corresponding to complete $pp$ calorimetry at the adopted acceleration normalization. Applying the same calculation to M82 and NGC~253 gives $\rho_{\rm starburst}=\rhostarburst$. The two are separated by a factor of $\rhoratio$ in the median, with a 68 percent interval of $\rhoratiolo$--$\rhoratiohi$. The ratio is formed realization by realization; the uncertainty on $\eta_{\max}$, which is common to both classes, therefore cancels exactly; what remains is the published uncertainty on the stacked wind efficiency and the $0.3$~dex systematic we assign to the superwind mechanical powers. The Monte Carlo then gives $P(\rho_{\rm starburst}>\rho_{\rm wind})=\Pgtone$ and $P(\rho_{\rm starburst}>3\,\rho_{\rm wind})=\Pgtthree$. The ordering is therefore secure and the factor of three is established at about the $2\sigma$ level; the exact size of the gap is not. Table~\ref{tab:calorimetry} summarizes the budget for all four classes.

\begin{table}
\caption{Calorimetric budget of mechanical outflows in the local Universe.}
\label{tab:calorimetry}
\centering
\scriptsize
\setlength{\tabcolsep}{1.5pt}
\begin{tabular}{lcccccc}
\hline\hline
Class & $N$ & $\dot{E}$ & $L_\gamma$ & $\eta$ & $\rho$ & $n_{\rm req}$ \\
 & & (erg s$^{-1}$) & (erg s$^{-1}$) & & & (cm$^{-3}$) \\
\hline
\multicolumn{7}{l}{\textit{Seyfert AGN winds (stacked)}} \\
\quad S-WIND\tablefootmark{a} & 11 & $1.8\times10^{44}$ & $1.0\times10^{41}$ & $5.2\times10^{-4}$ & 0.020$^{+0.018}_{-0.011}$ & $\sim7$ \\
\hline
\multicolumn{7}{l}{\textit{Starburst superwinds}} \\
\quad M82 & 1 & $3.1\times10^{42}$ & $1.5\times10^{40}$ & $5.0\times10^{-3}$ & 0.20 & \ldots \\
\quad NGC~253 & 1 & $1.5\times10^{42}$ & $1.3\times10^{40}$ & $8.8\times10^{-3}$ & 0.35 & \ldots \\
\quad combined & 2 & \ldots & \ldots & \ldots & 0.26$^{+0.37}_{-0.16}$ & $\sim50$ \\
\hline
\multicolumn{7}{l}{\textit{LINER ionized outflows: most constraining limit}} \\
\quad NGC~1052 & 1 & $8.8\times10^{40}$ & $<$$2.8\times10^{40}$ & $<$$3.2\times10^{-1}$ & $<13$ & \ldots \\
\hline
\multicolumn{7}{l}{\textit{Radio-galaxy jets (reference only; not hadronic calorimeters)}} \\
\quad M87 & 1 & $2.4\times10^{43}$ & $4.9\times10^{41}$ & $2.0\times10^{-2}$ & (0.80) & \ldots \\
\quad Cen~A & 1 & $1.0\times10^{43}$ & $3.3\times10^{39}$ & $3.3\times10^{-4}$ & (0.01) & \ldots \\
\hline
\end{tabular}
\tablefoot{$L_\gamma$ is the 0.1--100~GeV luminosity and $\eta=L_\gamma/\dot{E}$, which holds
row by row for the individual objects; $\rho=\eta/\eta_{\max}$ with
$\eta_{\max}=\xi_{\rm CR}f_{\pi^0}f_{\rm band}=2.5\%$ for
$\xi_{\rm CR}=0.1$ (Sect.~\ref{sec:framework}); absolute values of $\rho$ scale as
$\xi_{\rm CR}^{-1}$, ratios between rows do not. $n_{\rm req}$ is the ambient proton
density that reproduces the observed $\rho$ for the fiducial geometry of each class, quoted to
one significant figure because it inherits the model dependence of $R$ and $v_{\rm sh}$
(Sect.~\ref{sec:density}). Uncertainties on $\rho$ are the 68\% Monte Carlo intervals of
Appendix~\ref{app:mc} and include the uncertainty on $\eta_{\max}$, which is common to all
rows. Values of $\rho$ in parentheses are formal and carry no calorimetric meaning, because
jet emission is neither hadronic nor isotropic. Only classes with both a measured $\gamma$-ray luminosity and a measured mechanical power appear here: S-XRAY and S-PAL have no tabulated outflow energetics, and S-DET is tested instead against the star-formation ceiling in Sect.~\ref{sec:sfceiling}. Of the six LINERs only NGC~1052 is listed, since it is the one whose limit falls below $\eta=1$; the others are shown in Fig.~\ref{fig:master}.}
\tablefoottext{a}{$\eta$ for S-WIND is the stacked-fit efficiency measured directly by
\citet{Ajello2021}, band-corrected; it is not the ratio of the two tabulated sample means,
which is $5.7\times10^{-4}$, because the stacking fits a single efficiency to all eleven
sources instead of dividing sample averages. The 10 percent difference is smaller than every
uncertainty quoted here.}
\end{table}

One sensitivity deserves a note, because the kinetic powers are the weakest input. We adopt the efficiency measured by \citet{Ajello2021} rather than recombining the per-source powers ourselves; their quoted sample mean, $1.8\times10^{44}$~erg~s$^{-1}$, lies between the logarithmic and linear means of the published intervals. Resampling each $\dot{E}_{{\rm K},i}$ log-uniformly within its own minimum--maximum range and taking the linear sample mean instead would give $\rho_{\rm wind}=\rhoresampled$, lower by $\rhoshiftdex$~dex, because the linear mean is dominated by the upper ends of intervals up to $1.7$~dex wide; sampling uniformly in linear space instead lowers $\rho_{\rm wind}$ further, to $\rholinuniform$. In either case the contrast with starbursts exceeds three in more than $\pctcontrastgtthree$ percent of realizations. The direction matters more than the size: any such recombination makes Seyfert winds less calorimetric and widens the contrast with starbursts, which makes the value adopted here the conservative one. The opposite extreme is the one that matters. Placing every source at its tabulated maximum gives $\rho_{\rm wind}\simeq3\times10^{-4}$ and a contrast of $\contrastmax$, while placing every source at its tabulated minimum gives $\rho_{\rm wind}=\rhowindmin$ and a contrast of only $\contrastmin$, which would erase the result. That configuration is a bound, not a plausible realization --- for $\nlowerlim$ of the eleven sources the tabulated minimum is itself a lower limit, and it requires every wind in the sample to sit simultaneously at its smallest permitted launch radius --- but it defines the condition under which the central comparison fails, and we state it as such.

We emphasize that $\rho<1$ is not a criticism of the wind interpretation: it is what a sub-calorimetric shock is expected to produce, and it is the quantitative content of the wind hypothesis. The key point is that the ratio is small and therefore informative.

\subsection{What the calorimetry requires: a density contrast}
\label{sec:density}

\begin{figure*}[t]
\centering
\includegraphics[width=\textwidth]{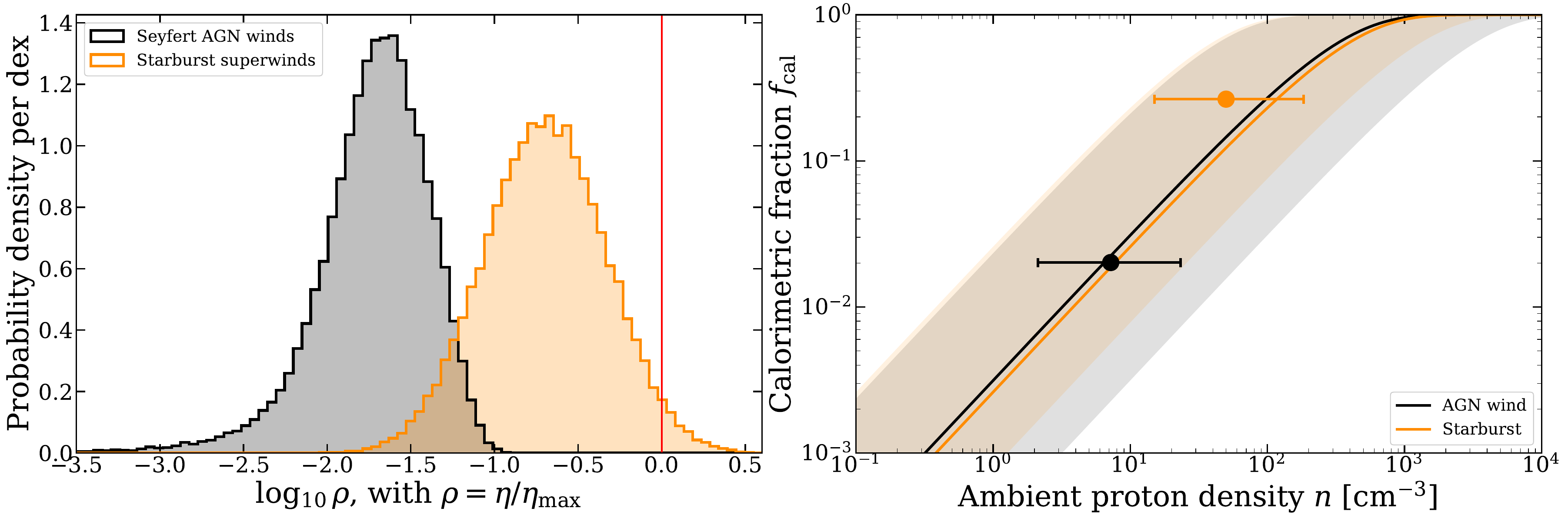}
\caption{Left: Monte Carlo posterior of the calorimetric ratio $\rho=\eta/\eta_{\max}$ for Seyfert AGN winds and for starburst superwinds, including the uncertainty on $\eta_{\max}$ (which is common to both). Dotted vertical lines mark the medians. Right: the proton calorimetric fraction of Eq.~(\ref{eq:fcal}) as a function of ambient density, for the fiducial geometry of each class (solid) and for the full ranges of $R$ and $v_{\rm sh}$ explored (shaded). Points with error bars show the density required to reproduce each observed $\rho$. The two curves nearly coincide by construction: similar radii and shock speeds are adopted for the two classes, and the difference in $\rho$ therefore maps directly onto a difference in density and not into a difference in geometry.}
\label{fig:calorimetry}
\end{figure*}

Inverting Eq.~(\ref{eq:fcal}) for the fiducial geometries gives the densities of Table~\ref{tab:calorimetry} and Fig.~\ref{fig:calorimetry}:
\begin{equation}
n_{\rm wind}\sim10\ {\rm cm^{-3}},\qquad
n_{\rm starburst}\sim10^{2}\ {\rm cm^{-3}}.
\end{equation}
These are densities required within the adopted transport geometry, not measured or uniquely inferred environmental densities. Because $f_{\rm cal}\ll1$ for both classes, Eq.~(\ref{eq:fcal}) is in its linear regime, $f_{\rm cal}\simeq t_{\rm res}/t_{pp}\propto nR/v_{\rm sh}$; density, size and shock speed thus enter only through that combination and are exactly degenerate. We therefore quote only order-of-magnitude values. The numerical 68 percent intervals (Table~\ref{tab:calorimetry} and Appendix~\ref{app:mc}) are dominated by the assumed ranges of $R$ and $v_{\rm sh}$, which are taken from one-dimensional models of a different sample, and the identification $\rho=f_{\rm cal}$ carries the caveats of Eq.~(\ref{eq:rhofcal}); quoting the density to two significant figures would therefore be misleading. The implied contrast is about one order of magnitude, with $P(n_{\rm starburst}>n_{\rm wind})=\Pcontrast$. Because the scaling $n_{\rm req}\propto v_{\rm sh}/R$ is exact in this regime, the effect of a more compact emission region follows directly: at $R=10$~pc the wind would require $n\simeq\nattenpc$~cm$^{-3}$, and the contrast with starbursts disappears for $R\lesssim\Rerase$~pc. The adopted $20$--$300$~pc comes from the UFO-bubble models cited above, and is an order of magnitude larger than the extended corona invoked by \citet{Liu2025}, which represents a different emission site, not a smaller-scale version of the same region; but a wind shock on parsec scales would remove the density contrast, and that assumption is what this part of the interpretation rests on.

These values are physically plausible. A few particles per cm$^3$ is what is expected of the diffuse, largely evacuated medium through which a wind bubble expands after a few $10^5$~yr in a galaxy bulge; a hundred particles per cm$^3$ is a conservative value for the molecular ISM of a starburst nucleus, where the CRs that produce the observed $\gamma$ rays are confined. If the two accelerator classes have comparable shock speeds and comparable sizes --- which is imposed, not measured --- then what differs is the target density, and that difference alone accounts for the order of magnitude in $\rho$. This is one possible interpretation of the contrast, but not a unique inference: the residence time need not be $R/v_{\rm sh}$ if escape is diffusive, and a systematic difference in residence time between the two classes would reproduce the same contrast without requiring any difference in density.

\subsection{The individually detected Seyferts do not need an AGN component}
\label{sec:sfceiling}

\begin{figure}[t]
\centering
\includegraphics[width=\columnwidth]{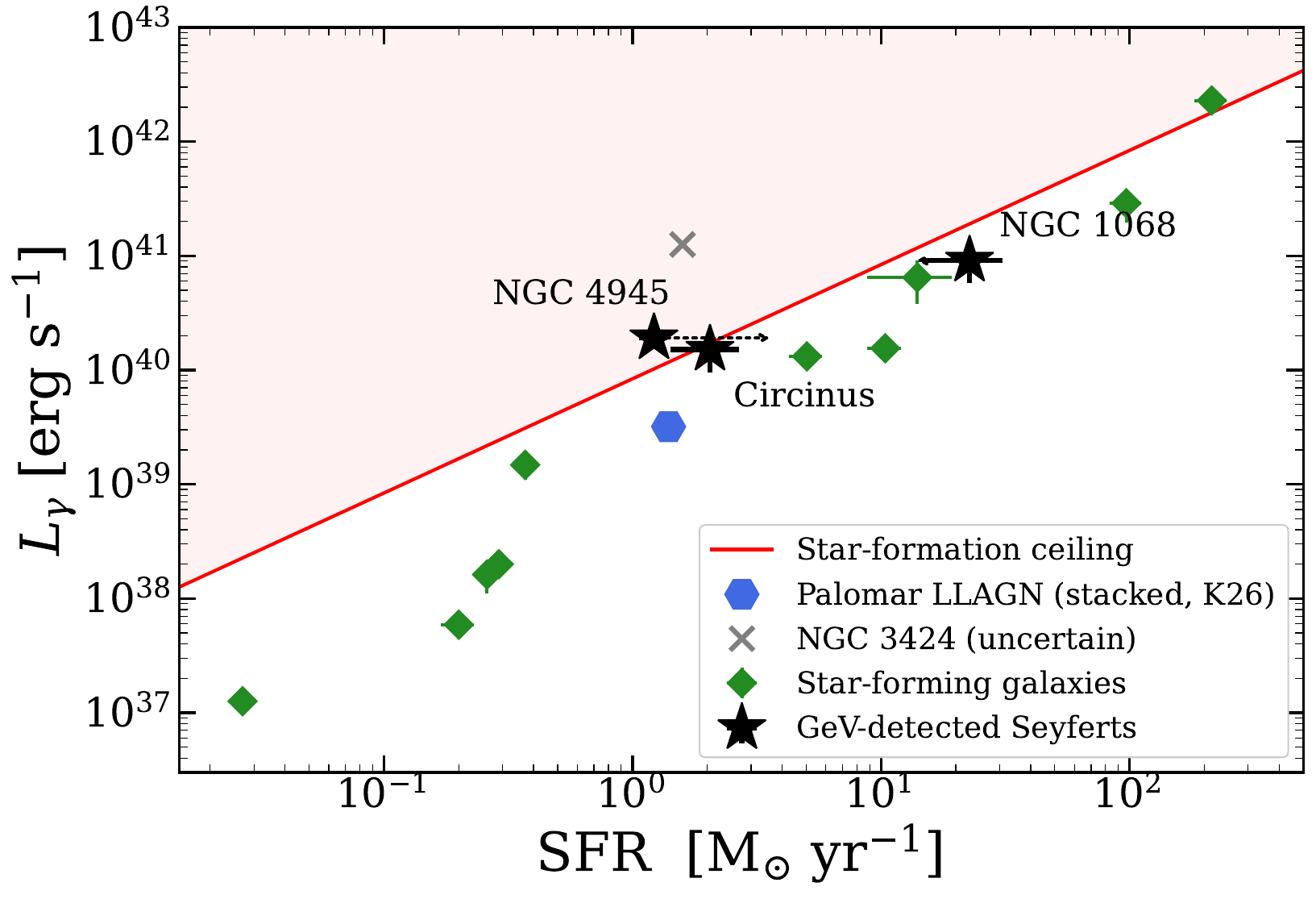}
\caption{The three individually GeV-detected radio-quiet Seyferts (stars) against the star-formation calorimetric ceiling (solid line; the shaded region lies above the ceiling), together with the star-forming galaxies of \citet{Kornecki2020}. Dotted arrows show where each Seyfert moves if the IR-based star-formation rate is used instead of the H$\alpha$+IR tracer. NGC~3424, whose AGN content is disputed, is shown separately. The hexagon is the stacked Palomar LLAGN measurement of \citet[][K26]{Karwin2026}, placed using the same infrared calibration of the star formation rate.}
\label{fig:sfcal}
\end{figure}

A hadronic calorimeter powered by star formation obeys the same ceiling, with the supernova mechanical power in place of $\dot{E}$; \citet{Kornecki2020} write it as $L_\gamma^{\max}=8.4\times10^{39}\,({\rm SFR}/M_\odot\,{\rm yr^{-1}})$~erg~s$^{-1}$, where the star formation rate (SFR) is in $M_\odot$~yr$^{-1}$. Figure~\ref{fig:sfcal} tests the three individually detected radio-quiet Seyferts against it. Using the H$\alpha$ plus IR star-formation rates of \citet{Kornecki2020}, NGC~1068 lies at a star-formation (SF) calorimetric ratio $\rho_{\rm SF}=0.48$ and Circinus at $0.88$, both below the ceiling, while NGC~4945 lies at $1.86$, apparently above it.

The apparent excess for NGC~4945 is not robust. \citet{Kornecki2020} themselves flag this galaxy as a $4.2\sigma$ outlier of the SFR--$L_{\rm IR}$ relation and attribute it to its strong obscuration and near-edge-on geometry, which cause the H$\alpha$+IR tracer to underestimate the true star-formation rate. Adopting instead the IR-based rate, ${\rm SFR}=1.7\times10^{-10}\epsilon\,L_{\rm IR}/L_\odot$ with $\epsilon=0.79$ for a Chabrier initial mass function, moves NGC~4945 to $\rho_{\rm SF}=0.61$ and, propagating the published uncertainties, the probability of exceeding the ceiling falls from $1.00$ to $0.02$. The conclusion depends entirely on the choice of star-formation tracer, and we therefore do not claim a detection of an AGN-powered component in any of the three. The robust conclusion is instead the negative one: the GeV emission of every radio-quiet Seyfert in S-DET is consistent with being powered by star formation alone.

This conclusion does not necessarily extend to all radio-quiet Seyferts. S-DET is not a complete sample of the class but the set of individually detected objects that appear in one homogeneous compilation, and the ceiling test cannot be applied to the stacked populations at all without assuming an average star formation rate for them. \citet{LiuJR2025} report NGC~3281 at $6.22\sigma$ and argue that its star formation is too weak to supply the observed luminosity, favouring inverse-Compton emission from the corona or a weak jet; the source is Compton-thick, with an estimated $\gamma\gamma$ optical depth of order ten, which implies an even larger intrinsic luminosity. NGC~3281 is not in S-DET because it does not appear in the homogeneous compilation of \citet{Kornecki2020}, which leaves no star formation rate on the same distance scale available for it, and because \citet{Liu2025} excluded it from their own sample on the grounds that it lies close to a region of extended residual emission. It nevertheless provides the clearest counterexample to a purely star-formation origin for individually detected radio-quiet Seyferts, and it favors a coronal origin over a wind origin. Lying below the ceiling means that star formation is energetically capable of supplying the observed luminosity within the adopted calibration; it neither demonstrates that the $\gamma$ rays originate there nor sets a strong limit on an additional AGN component.

\section{One population or two?}
\label{sec:two}

\subsection{Can the two stacking detections be distinguished?}

Before comparing the two stacks it is necessary to establish whether their host samples can be treated as draws from a single parent population. They cannot be treated as such in full. A two-sample Kolmogorov--Smirnov test finds the black-hole mass distributions compatible ($D=0.26$, $p=0.52$; medians $\log M_\bullet=7.60$ and $7.39$), but the distance distributions incompatible ($D=0.55$, $p=0.007$; medians $61$ and $37$~Mpc) and the bolometric luminosities incompatible as well ($D=0.57$, $p=0.005$ for $\kappa_{14-195}=8$; medians $\log L_{\rm bol}=44.4$ and $43.9$). This difference is unsurprising --- one sample is selected on UFO detection out to $z<0.1$, the other on hard X-ray flux within $60$~Mpc --- but it means that the comparison below involves two differently selected samples, and that only the black-hole mass axis is matched. We therefore normalize by $L_{\rm Edd}$, and we treat the exercise as a null test rather than as a measurement of a common efficiency.

\begin{figure*}[t]
\centering
\includegraphics[width=\textwidth]{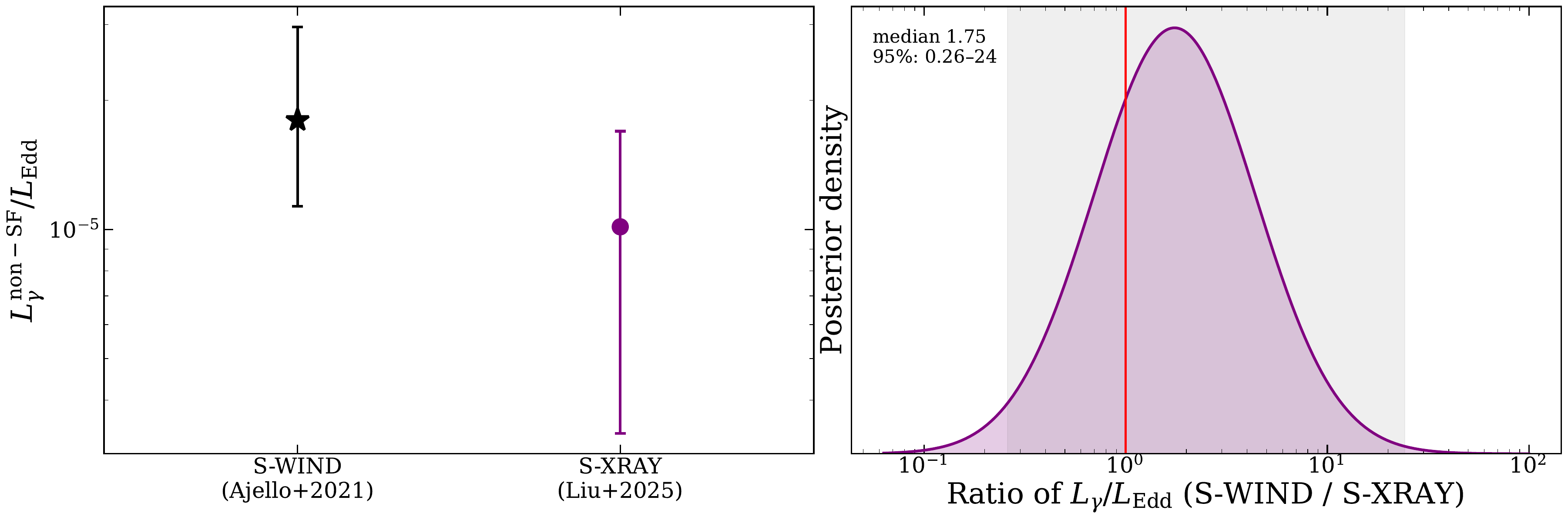}
\caption{Left: star-formation-subtracted $\gamma$-ray luminosity per unit Eddington luminosity for the two stacked Seyfert samples. Error bars are the published asymmetric luminosity uncertainties ($+5.1/-2.9$ and $\pm1.0$ on $7.9$ and $1.5\times10^{40}$~erg~s$^{-1}$ respectively) propagated through the band conversion and the star-formation subtraction. Right: Monte Carlo posterior for their ratio; the dashed line marks equality and the shaded band the 95\% interval.}
\label{fig:twostacks}
\end{figure*}

Both stacking analyses estimate the star-formation contribution to their own signal and find it small. Using the \citet{Kornecki2020} relation, which gives the larger and therefore more conservative estimate, star formation accounts for $11$ percent of the S-WIND stack and $16$ percent of the S-XRAY stack; using the \citet{Ajello2020} relation, as both original papers did, the fractions are $4$ and $6$ percent. The choice has little effect: the ratio computed below is $1.77$ with the \citet{Kornecki2020} subtraction, $1.72$ with the \citet{Ajello2020} subtraction and $1.69$ with no subtraction at all, because the correction is small and acts in the same direction in both samples. We adopt the conservative estimate, with its $0.34$~dex calibration scatter. That scatter is carried through the Monte Carlo, but it is not what sets the width of the posterior in Fig.~\ref{fig:twostacks}: switching it off entirely narrows the 95 percent interval from $1.96$ to $1.86$~dex, moves the median by less than $1$ percent, and changes $P(\mathrm{ratio}<1)$ from $0.256$ to $0.246$. Even doubling it to $0.68$~dex widens the interval by only a further $0.1$~dex. The tails are therefore dominated by the published uncertainties on the luminosities of the two stacked measurements, not by the calibration of the star-formation calorimeter (Appendix~\ref{app:mc}). The S-XRAY infrared luminosity, $\log(L_{\rm IR}/L_\odot)=10.12$, is read from the average spectral energy distribution of \citet{Liu2025}, not from a tabulated measurement. Since $L_\gamma\propto L_{\rm IR}^{1.21}$ is convex, evaluating it at the mean $L_{\rm IR}$ instead of averaging per-source predictions understates the star-formation contribution, by $\jensen$ percent for a $0.4$~dex spread in $L_{\rm IR}$; that would raise the subtracted fraction from $\sfxray$ to $17$ percent, which is negligible against the uncertainties that follow.

The comparison itself requires a normalization available for both samples without conversion, and the Eddington luminosity is the only such quantity: both papers tabulate black-hole masses. We stress that this is a convenience and not a physical expectation --- neither wind power nor coronal luminosity should scale with $M_\bullet$ at fixed accretion rate --- and we therefore do not treat the resulting number as an efficiency. We obtain
\begin{equation}
\frac{L_\gamma^{\rm non\text{-}SF}}{L_{\rm Edd}}=1.8\times10^{-5}\ \ ({\rm S\text{-}WIND}),\qquad
1.0\times10^{-5}\ \ ({\rm S\text{-}XRAY}),
\end{equation}
a ratio of $1.75$, with a 68 percent interval of $0.75$--$4.8$ and a 95 percent interval of $0.26$--$24$ (Fig.~\ref{fig:twostacks}). An interval spanning two orders of magnitude is compatible with equality but also with a factor-of-ten difference in either direction. The present data therefore have essentially no power to separate the two measurements. This is a limitation of the data, not a statement about the physics, and no physical claim is built on it. The width is dominated by the published luminosity uncertainties, which are themselves $\sim65$ percent in each case. We therefore treat this subsection as secondary to the calorimetric result: it is a useful null comparison, and the factor of $\twostackratio$ should not be read as evidence that the two mechanisms produce comparable $\gamma$-ray output.

\begin{figure}[t]
\centering
\includegraphics[width=\columnwidth]{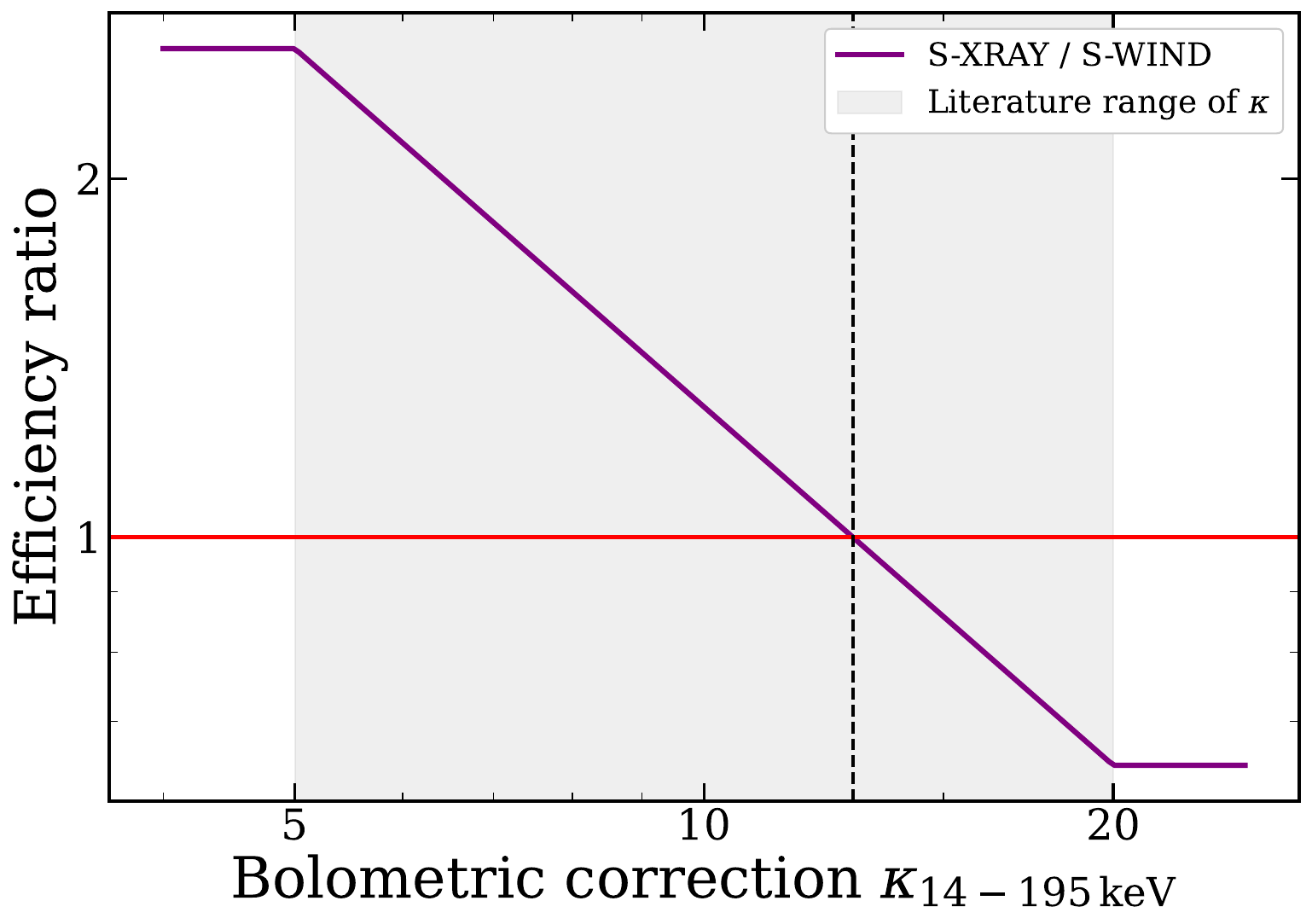}
\caption{Ratio of the star-formation-subtracted $\gamma$-ray efficiency of the two stacked Seyfert samples, in terms of AGN bolometric luminosity, as a function of the hard X-ray bolometric correction assumed for the S-XRAY sample. The shaded band is the range of $\kappa_{14-195}$ reported for local Seyferts; the dashed line marks equality.}
\label{fig:kappa}
\end{figure}

An alternative normalization is therefore worth having, and the natural one is the AGN bolometric luminosity, to which both a wind power and a coronal luminosity should plausibly scale. The same exercise can be done in those terms, but it then requires a hard X-ray bolometric correction for the S-XRAY sample, which is not tabulated. Rather than adopt one value, Fig.~\ref{fig:kappa} shows the ratio as an explicit function of $\kappa_{14-195}$: it falls from $2.6$ at $\kappa=5$ to $0.64$ at $\kappa=20$, crossing unity at $\kappa\simeq13$, a value within the range reported for local Seyferts \citep{VasudevanFabian2009, Vasudevan2010}. Because the ratio can cross unity within the plausible range of $\kappa$, the bolometric comparison carries no information beyond the $L_{\rm Edd}$ one: there the spread is set by the unknown bolometric correction rather than by the published luminosity uncertainties, and no choice of $\kappa$ in that range creates a discrepancy.

\subsection{Do the two pictures describe the same region?}

Consistency of the luminosities alone does not identify the emitting region, but the two independently inferred size scales are closer than might be inferred from the two interpretations. The extended corona invoked by \citet{Liu2025} to avoid $\gamma\gamma$ absorption above a few GeV has $R_{\rm ext}\simeq2.7\times10^6\,R_{\rm g}\simeq2.8$~pc for their mean black-hole mass; the wind forward shock modeled by \citet{Ajello2021} has traveled $20$--$300$~pc. These scales differ by one to two orders of magnitude --- a genuine difference, but substantially smaller than that between a canonical $10\,R_{\rm g}$ corona and a galaxy-scale wind, and both lie in the circumnuclear region, not in the accretion flow proper. \citet{Liu2025} note that their extended corona coincides with the $1$--$23$~pc scales of the 100~GHz emission resolved by the Atacama Large Millimeter/submillimeter Array in radio-quiet AGN, and also that a nuclear star cluster of $\lesssim10$~pc would produce emission on the same scale.

We do not claim that the two detections measure the same photons; the samples are largely disjoint and the two photon indices ($\Gamma=2.1\pm0.3$ and $2.32^{+0.40}_{-0.35}$) agree within their uncertainties and therefore provide no discrimination either. We do find that the case for two distinct mechanisms is weaker than suggested by the two papers considered separately: the luminosities agree, the inferred scales are within one to two decades of each other, and the wind hypothesis was not tested for the S-XRAY sample nor the coronal hypothesis for S-WIND.

The two samples are also not strictly independent: as noted in Sect.~\ref{sec:samples}, MCG--5-23-16 and NGC~4507 appear in both. The two sources contribute $TS=0.56$ of the S-XRAY total of $30.6$, i.e. $1.8$ percent; the corresponding change in the stacked flux is therefore expected to be negligible and neither detection depends on them. The relevant consequence is that for most of the remaining S-XRAY sources the absence of a detected UFO reflects the absence of a dedicated Fe~K search rather than a demonstrated absence of a wind; NGC~7172, which was searched and found to have none, is the exception that shows the qualification is needed.

\subsection{A source-to-source test, and a target}
\label{sec:targets}

\begin{figure}[t]
\centering
\includegraphics[width=\columnwidth]{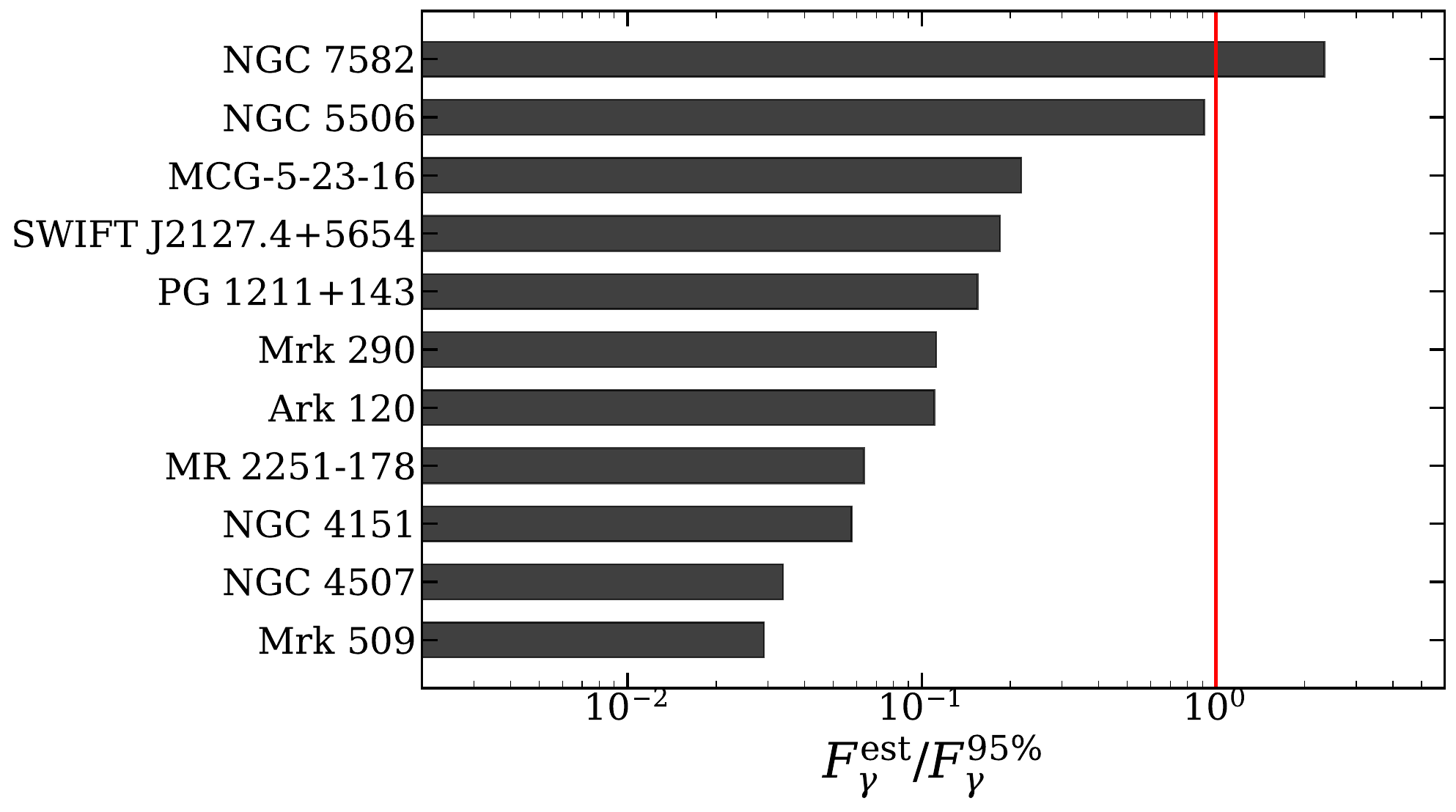}
\caption{Estimated wind-powered 0.1--100~GeV flux, obtained by applying the population-average efficiency to each source's own kinetic power, divided by that source's published 95\% confidence upper limit. Values above unity (dashed line) are already excluded. These are estimates under the assumption of a universal wind efficiency, not measurements of the individual sources.}
\label{fig:targets}
\end{figure}

If a single efficiency characterized every wind host, the population-average $\eta_{\rm wind}$ applied to each object's own kinetic power and distance would give a flux well below its own published upper limit. This is not the case. For NGC~7582 the predicted 0.1--100~GeV flux is $2.3$ times its own 95 percent limit (Fig.~\ref{fig:targets} and Table~\ref{tab:swind}), and NGC~5506 is at $0.9$. This test relies on two assumptions. First, we model the individual $\rho_i$ as log-normally distributed with their mean fixed to the stacking result; the published efficiency is consistent with a ratio of sample means, which constrains the $\dot{E}$-weighted mean of $\rho$ and coincides with the arithmetic mean only if $\rho$ is uncorrelated with $\dot{E}$ (Appendix~\ref{app:weights}). Second, we treat each published 95 percent limit as a hard threshold, which a soft-likelihood test shows to be the conservative choice (Appendix~\ref{app:censor}). Treating the eleven limits as censored data and requiring a log-normal $\rho$ distribution whose mean reproduces the stacking result, the limits demand a dispersion $\sigma_{\log\rho}>\sigmarho$~dex at 95 percent confidence, where the dispersion is that of the total scatter in $\rho$. Because each $\dot{E}_{{\rm K},i}$ carries a $0.6$--$1.7$~dex uncertainty, a substantial part of that scatter is measurement error, and the bound is therefore not by itself evidence for an intrinsic, density-driven variation in $f_{\rm cal}$ (the corresponding 68 percent bound is $>1.0$~dex; Appendix~\ref{app:mc}). We quote the weaker, more defensible bound.

Scatter of this size is consistent with the calorimetric picture, in which $f_{\rm cal}\propto n$ in the sub-calorimetric regime while circumnuclear densities range over orders of magnitude. It does not identify density as the cause: acceleration efficiency, shock speed and radius, residence time, wind duty cycle, covering factor, multiphase structure, nuclear star formation and weak-jet contamination would all produce source-to-source variation in the effective $\rho$. This bound is conditional --- it holds if $\rho$ is uncorrelated or anti-correlated with $\dot{E}_{\rm K}$, and disappears if the two are strongly positively correlated (Appendix~\ref{app:weights}). It also implies that the stacked signal is not uniformly shared across the sample, and that searches for individual detections should be targeted at the sources with the largest $\dot{E}_{\rm K}n/d^2$, and not simply those with the largest $\dot{E}_{\rm K}$.

Table~\ref{tab:swind} ranks the S-WIND sample accordingly. The three most favorable targets are NGC~7582, NGC~5506 and MCG--5-23-16; all are nearby ($D<40$~Mpc) Seyfert~2 or 1.9 systems, and all three are exactly the objects in which a dense circumnuclear medium is expected. NGC~7582 in particular is already constrained below the population-average efficiency. This result should not be over-interpreted. The kinetic power of NGC~7582 has a published minimum--maximum range of $1.5$~dex, and a factor of $2.3$ lies well inside it; the limit was derived over $11.1$~yr with $\Gamma$ fixed to $2.0$ while the prediction assumes the stacked $\Gamma=2.1$; and with eleven sources one $\sim2\sigma$-level outlier is unremarkable. The robust content of the test is therefore the censored-likelihood bound, $\sigma_{\log\rho}>0.3$~dex, and not the individual object.

\subsection{A third stacked measurement, and the weight of NGC 4151}
\label{sec:third}

The Palomar LLAGN stack of \citet{Karwin2026} is a third, independent measurement of the average GeV output of local Seyfert-like nuclei, and the only one of the three reaching below $L_{\rm bol}\sim10^{42}$~erg~s$^{-1}$. Their fitted $L_\gamma$--$L_{\rm IR}$ relation supplies a luminosity at the sample mean infrared luminosity without any distance assumption of ours; converted to the reference band it gives $L_\gamma\simeq\Lpal$~erg~s$^{-1}$. No outflow energetics are tabulated for these objects, and S-PAL therefore cannot be placed on the $L_\gamma$--$\dot{E}$ plane of Fig.~\ref{fig:master}; it can, however, be placed on the star-formation plane of Fig.~\ref{fig:sfcal}, where the same infrared luminosity converts to a rate of $1.4\,M_\odot$~yr$^{-1}$ and the stack falls a factor of four below the ceiling, consistent with the star-formation origin its authors infer. The three stacked measurements are then ordered as
\begin{equation}
L_\gamma^{\rm S\text{-}PAL}:L_\gamma^{\rm S\text{-}XRAY}:L_\gamma^{\rm S\text{-}WIND}\simeq1:\ratioFGRPAL:\ratioUFOPAL,
\end{equation}
spanning $\spandex$~dex, in the same order as the AGN power of the three selections: optically selected low-luminosity nuclei, hard-X-ray-bright Seyferts, and UFO hosts. This ordering should not be interpreted as a measurement of a common efficiency --- the three samples differ in selection, distance and normalization, and only the two of Sect.~\ref{sec:two} can be placed on a common footing at all. It is nevertheless the behavior expected if the GeV output of a non-jetted Seyfert tracks its accretion power.

Two further results of \citet{Karwin2026} bear on our interpretation. Their stacked Seyfert subset ($N=40$, $TS=11.0$) and LINER subset ($N=84$, $TS=5.2$) have spectral indices consistent within $1\sigma$, with the Seyfert flux about twice the LINER flux; this is the first population-level $\gamma$-ray comparison of the two classes. At $TS=5.2$, however, the LINER index is only weakly constrained, and this agreement should not be read as evidence that the two classes share spectral properties. And their $L_\gamma$--$L_{\rm IR}$ slope, $\alpha=1.4\pm0.35$, agrees with the star-forming-galaxy relation within $1\sigma$, independently supporting the conclusion of Sect.~\ref{sec:sfceiling} that star formation suffices to explain the emission of the individually detected radio-quiet Seyferts.

Finally, the contribution of NGC~4151 to the stacked signal deserves quantification. \citet{Ajello2021} publish an alternative stack with that source removed, with $TS=\TSnoNfourone$ instead of $30.1$ and a reference-band energy flux $0.82$ times the nominal one. Propagating that factor gives $\rho_{\rm wind}=\rhowindnoNfourone$ in place of $\rhowind$ --- a shift well inside the 68 percent interval, and one that makes Seyfert winds slightly less calorimetric, not more. The central comparison of this paper is therefore robust to excluding the one source whose association is contested, and the direction of the shift strengthens rather than weakens the conclusion. There is also an internal tension worth recording. NGC~4151 ranks ninth of eleven in Table~\ref{tab:swind}: its kinetic power is the lowest in the sample, and at the population-average efficiency it would be expected to be among the faintest wind emitters, yet it carries half the stacked significance. If its contribution were genuinely wind-powered, its calorimetric ratio would have to be $\rho\simeq0.35$, seventeen times the population average and above the starburst value. That is not impossible, but it is either a large part of the dispersion required in Sect.~\ref{sec:targets} or an indication that the emission attributed to NGC~4151 is not wind-powered at all.

\section{Discussion}
\label{sec:discussion}

\subsection{Power, calorimetry, and why LINERs and Seyferts fail differently}

Taken together, the LINER and Seyfert results provide a compact picture of what determines GeV detectability in non-jetted AGN. The detectable luminosity is the product of two independent factors,
\begin{equation}
L_\gamma\simeq\eta_{\max}\,f_{\rm cal}(n,R,v_{\rm sh})\;\dot{E},
\label{eq:product}
\end{equation}
and the local AGN population is limited by one factor or the other. LINER ionized outflows have $\dot{E}\sim10^{38}$--$10^{41}$~erg~s$^{-1}$ \citep{HermosaMunoz2024, Cazzoli2022}: even at starburst-like calorimetry they would fall three to four orders of magnitude below current thresholds, which is precisely the conclusion of D26. Seyfert winds have $\dot{E}\sim10^{44}$~erg~s$^{-1}$, three decades more, but $f_{\rm cal}\sim0.02$; they thus emit only $\sim10^{41}$~erg~s$^{-1}$ and remain individually undetectable. Starburst superwinds have intermediate power, $\sim10^{42}$~erg~s$^{-1}$, but an order of magnitude better calorimetry, and are detected. Relativistic jets fall outside this framework: they are neither hadronic calorimeters nor isotropic, which is why M87 sits near the nominal ceiling in Fig.~\ref{fig:master} without any of this reasoning applying to it.

There is now a population-level $\gamma$-ray constraint on LINERs to set beside the six resolved outflows of D26: the 84 Palomar LINERs stacked by \citet{Karwin2026} give only $TS=5.2$ ($1.8\sigma$), and the 66 without broad H$\alpha$ are consistent with zero. Neither the resolved nor the stacked approach detects the class, which is what Eq.~(\ref{eq:product}) predicts: at $\dot{E}\sim10^{38}$--$10^{41}$~erg~s$^{-1}$, no plausible calorimetry reaches the LAT threshold.

This also clarifies the role of the compact jet in the LINER result. D26 concluded that efficient $\gamma$-ray LINERs such as NGC~4278 require a compact nuclear jet because their outflows cannot supply the power. The present analysis shows the complementary point for Seyferts: even when the power is available, in quantities exceeding those of any starburst, an extended AGN wind cannot convert it efficiently. Neither population reaches detectability through its ionized or fast outflow alone.

\subsection{Competing mechanisms}

Our framework constrains the hadronic wind channel but does not exclude the alternative mechanisms, which are summarized below.

Coronae. Leptonic inverse-Compton emission from non-thermal electrons in a compact corona \citep[e.g.,][]{Inoue2019, Inoue2021, Gutierrez2021} is not subject to Eq.~(\ref{eq:ceiling}), and the $\gamma\gamma$ opacity argument of \citet{Liu2025} is independent of anything computed here. Our result is that the wind channel can produce a luminosity of the observed order, and a coronal origin is therefore not required by the energetics; distinguishing them needs spectral or spatial information, not more efficiency arguments.

Low-power jets. Both source papers argue against jets from the $L_{\rm 1.4\,GHz}/L_{14-195\,\rm keV}\sim10^{-5}$ ratio characteristic of coronal rather than jet radio emission \citep[e.g.,][]{LaorBehar2008, Smith2020}, and from the fact that the radio-matched control samples are undetected. We do not further constrain this interpretation, except to note that the same radio-quiet selection was applied to both samples, and this argument cannot distinguish between them.

Star formation. Subdominant, at the $5$--$16$ percent level, in both stacks (Sect.~\ref{sec:two}); dominant, or at least sufficient, in all three individually detected systems (Sect.~\ref{sec:sfceiling}).

Molecular outflows. A stacking analysis of galaxies hosting molecular outflows \citep{McDaniel2023} probes the same shock physics in a colder, denser phase, where Eq.~(\ref{eq:product}) predicts a higher $f_{\rm cal}$ at the same mechanical power. Combining the molecular and ionized phases into a single multiphase energy budget would be a natural extension, but is beyond what published tables presently allow.

Proton--photon interactions. In the dense radiation field of a Seyfert nucleus $p\gamma$ processes can compete with $pp$ \citep[e.g.,][]{Inoue2019, Murase2024}, and would add a channel not captured by Eq.~(\ref{eq:fcal}). This would increase $f_{\rm cal}$ at fixed $n$ and therefore reduce the inferred density; the effect is strongest at small radii and is negligible at the $20$--$300$~pc scales relevant to the wind shock, but it is a caveat on the density and not on the calorimetric ratio.

\subsection{What would falsify this}
\label{sec:falsify}

The picture makes three predictions that current or near-future data can test. First, $\rho$ should correlate with an independent tracer of circumnuclear density --- CO or HCN surface density, or the [S~II]-derived electron density of the ionized gas --- across the wind sample; a stacking analysis binned by molecular gas surface density instead of $L_{\rm bol}$ would be the direct test. A signal with the expected sign is already present: \citet{Karwin2026} find that essentially all of the LLAGN stacked signal comes from spiral hosts ($6.0\sigma$ for 107 spirals against $0.8\sigma$ for 79 non-spirals), and spirals are precisely the systems with the dense circumnuclear gas that Eq.~(\ref{eq:product}) requires. We are careful not to claim this as confirmation: gas content and star formation rate are strongly correlated, and \citet{Karwin2026} themselves attribute the split to star-formation activity. This degeneracy is precisely why the decisive test should bin by gas density at fixed star formation rate, and not by morphology. With 11 and 37 sources respectively, two bins are realistically the maximum for the present samples; a decisive version of the test requires the full hard-X-ray-selected parent sample. Second, as LAT exposure accumulates the first individual detections should come from the top of Table~\ref{tab:swind}; NGC~7582 is already constrained below the population-average efficiency; once detected, it should appear at a flux below the population-average prediction, not at it. Third, because $pp$ emission extends to TeV with no attenuation at these scales whereas coronal emission is cut off by $\gamma\gamma$ absorption, the Cherenkov Telescope Array Observatory (CTAO) would see the wind component of the nearest, densest systems but not the coronal one. The required sensitivity can be stated, as an illustration and not as a prediction of the same evidentiary standing as the archival constraints, since it extrapolates a GeV spectrum by two decades: extrapolating the population-average wind spectrum with $\Gamma=2.1$ and no attenuation, and applying the EBL optical depths of \citet{SaldanaLopez2021}, the two best targets give $F(1-10\,{\rm TeV})\simeq\Fteva$ (NGC~7582, $22$~Mpc) and $\simeq\Ftevb$~erg~cm$^{-2}$~s$^{-1}$ (NGC~5506, $26$~Mpc); the EBL suppresses the band flux by $\eblsupp$ percent. These are at the level of the deepest projected CTAO point-source sensitivities: such a measurement would be challenging but feasible; a joint LAT--CTAO spectrum could distinguish between the two hypotheses in a way that efficiency arguments alone cannot. \citet{Sakai2026} reach the same targets from the opposite direction: calibrating a lepto-hadronic disk-wind model on nearby LAT-detected Seyferts, they identify UFO hosts as the most promising objects for TeV $\gamma$-ray and TeV--PeV neutrino observations. Their population-synthesis result, that disk winds supply at most ${\sim}5$ percent of the cosmic $\gamma$-ray background above $10$~GeV and ${\sim}10$ percent of the neutrino background near $100$~TeV, is an independent statement of the same inefficiency measured here, obtained without reference to a calorimetric ratio.

\subsection{Limitations}

Four limitations merit emphasis. (i) We perform no new LAT analysis: our $\gamma$-ray inputs are published stacked measurements and published per-source limits, obtained with different exposures (11.1 and 15~yr), different catalog versions and slightly different pipelines. A single homogeneous re-analysis of both samples using the 18~yr of data now available, following the ten-component setup of D26, would remove this heterogeneity. (ii) The kinetic powers of the S-WIND sample are uncertain by $0.6$--$1.7$~dex, because the launch radius of a UFO is not measured; this is the dominant systematic on $\eta$ and hence on $\rho$, and it is propagated throughout. It is also the quantity that Sect.~\ref{sec:rho} identifies as the one on which the central contrast depends, and the wind energetics now being measured with the X-Ray Imaging and Spectroscopy Mission (XRISM) \citep[e.g.,][]{Sakai2026} are the observations most likely to settle it. (iii) The densities of Sect.~\ref{sec:density} depend on shock radii and speeds taken from one-dimensional models, and should therefore be interpreted as order-of-magnitude estimates. (iv) The association of the brightest S-WIND source, NGC~4151, is contested \citep{Peretti2025, Karwin2026}; removing it lowers $\rho_{\rm wind}$ by 18 percent, which does not change any conclusion (Sect.~\ref{sec:third}), but it is the single source on which the S-WIND stacked significance most depends. The calorimetric ratios do not depend on the assumed shock radius and speed, since $\rho$ is defined directly as $\eta/\eta_{\max}$; they nevertheless depend on the mechanical-power estimates and the published $\gamma$-ray measurements, i.e.\ on (i) and (ii).

\section{Summary and conclusions}
\label{sec:conclusions}

We have placed the local AGN population on a common calorimetric footing, combining five populations: the three \textit{Fermi}-LAT stacking measurements S-WIND, S-XRAY and S-PAL, the individually GeV-detected radio-quiet Seyferts of S-DET, and our own LINER constraints of L-OUT, calibrated throughout against the star-forming-galaxy calorimeter. Our conclusions are:

\begin{enumerate}
\item For $\xi_{\rm CR}=0.1$, Seyfert AGN winds radiate $\rho=\rhowind$ of the $\gamma$-ray luminosity allowed by their mechanical power, a factor of $\simeq\rhoratio$ below starburst superwinds ($\rho=\rhostarburst$); the difference exceeds a factor of three with probability $\Pgtthree$. This result is conditional on the adopted wind energetics: the kinetic powers carry uncertainties of $0.6$--$1.7$~dex, and the contrast would be erased if every wind sat simultaneously at its minimum tabulated kinetic power (Sect.~\ref{sec:rho}). Removing NGC~4151, whose association is contested and which dominates the original stacked significance, lowers $\rho_{\rm wind}$ by $18$ percent and widens the contrast. Unlike the absolute values, the ratio is independent of the calorimetric normalization when the same parameters are adopted for both classes, and allowing class-specific acceleration and spectral parameters leaves the contrast robust (Appendix~\ref{app:xidiff}).
\item For the fiducial shock geometry this is consistent with the winds expanding into a medium of order $\nwind$~cm$^{-3}$, against order $\nstarburst$~cm$^{-3}$ for superwinds. These are consistency values, not measurements: density, shock radius and shock speed enter only through the combination $nR/v_{\rm sh}$ and are exactly degenerate. With similar shock speeds and sizes assigned to the two classes, the contrast is consistent with a lower target density in the Seyfert-wind environment, but an equivalent difference in residence time would reproduce it without any difference in density.
\item The two stacking detections, attributed in the literature to winds and to coronae, give the same star-formation-subtracted $\gamma$-ray output per unit Eddington luminosity to within a factor of $\twostackratio$, but with a 95 percent interval of $\twostacklo$--$\twostackhi$: the data cannot separate them, and the two samples are not matched in distance or AGN luminosity. The samples are also not strictly disjoint --- MCG--5-23-16 and NGC~4507 belong to both, contributing $1.8$ percent of one stacked test statistic.
\item None of the three radio-quiet Seyferts in S-DET exceeds the star-formation calorimetric ceiling once the systematic uncertainty of the star-formation-rate tracer is included; NGC~3281, detected more recently and not in that sample, is a counterexample whose emission star formation cannot supply \citep{LiuJR2025}; the apparent excess of NGC~4945 disappears with an IR-based rate.
\item A third stacked measurement, of 186 subthreshold Palomar LLAGN \citep{Karwin2026}, extends the comparison to much lower AGN power: the three stacked Seyfert-like populations are ordered $1:\ratioFGRPAL:\ratioUFOPAL$ in $\gamma$-ray luminosity, spanning $\spandex$~dex in the same order as the AGN power of their selections. Their $L_\gamma$--$L_{\rm IR}$ slope agrees with the star-forming-galaxy relation, and their signal comes almost entirely from spiral hosts --- consistent with a density-driven calorimetry, although this interpretation is degenerate with star formation.

\item The published per-source limits, treated as censored data, require $\sigma_{\log\rho}>\sigmarho$~dex (95 percent confidence) of source-to-source dispersion in the calorimetric ratio, provided that ratio is uncorrelated with, or anti-correlated with, the kinetic power; the dispersion is consistent with, though it does not establish, a density-driven variation in $f_{\rm cal}$. NGC~7582, whose estimated flux exceeds its own limit by a factor of $2.3$, largely drives this bound but is individually consistent with its large kinetic-power uncertainty.
\end{enumerate}

The unifying statement is that GeV detectability of a non-jetted AGN is set by the product of mechanical power and calorimetric fraction, evaluated on the spatial scale where the $\gamma$ rays are produced. For the Seyferts that scale is assumed, not directly measured: the mechanical power is inferred from parsec-scale UFO properties, while the emission is attributed to a shock two to three decades larger. With that caveat: LINER outflows have the calorimetry but not the power; the UFO-selected Seyfert winds have the power but not the calorimetry; individual detection becomes possible only when both are favorable, as in starburst--AGN composites. Testing this requires stacking binned by circumnuclear gas density instead of AGN luminosity, and TeV observations of the nearest, densest wind hosts.

\begin{acknowledgements}
AD acknowledges support from MCIN/AEI under grant PID2025-171816NB-I00 (MCIN/AEI/10.13039/501100011033/FEDER, EU). We thank the anonymous referee for a detailed and constructive report that substantially improved this paper. The \textit{Fermi} LAT Collaboration acknowledges generous ongoing support from a number of agencies and institutes that have supported both the development and the operation of the LAT as well as scientific data analysis.
\end{acknowledgements}

\section*{Data availability}
\label{sec:data}

This paper uses no proprietary data. Every input is a published measurement, referenced in Tables~\ref{tab:swind} and \ref{tab:calorimetry} and in the works cited there.

\appendix
\nolinenumbers

\section{Monte Carlo propagation and censored statistics}
\label{app:mc}

\subsection{Priors and procedure}

\begin{figure}[h]
\centering
\includegraphics[width=\columnwidth]{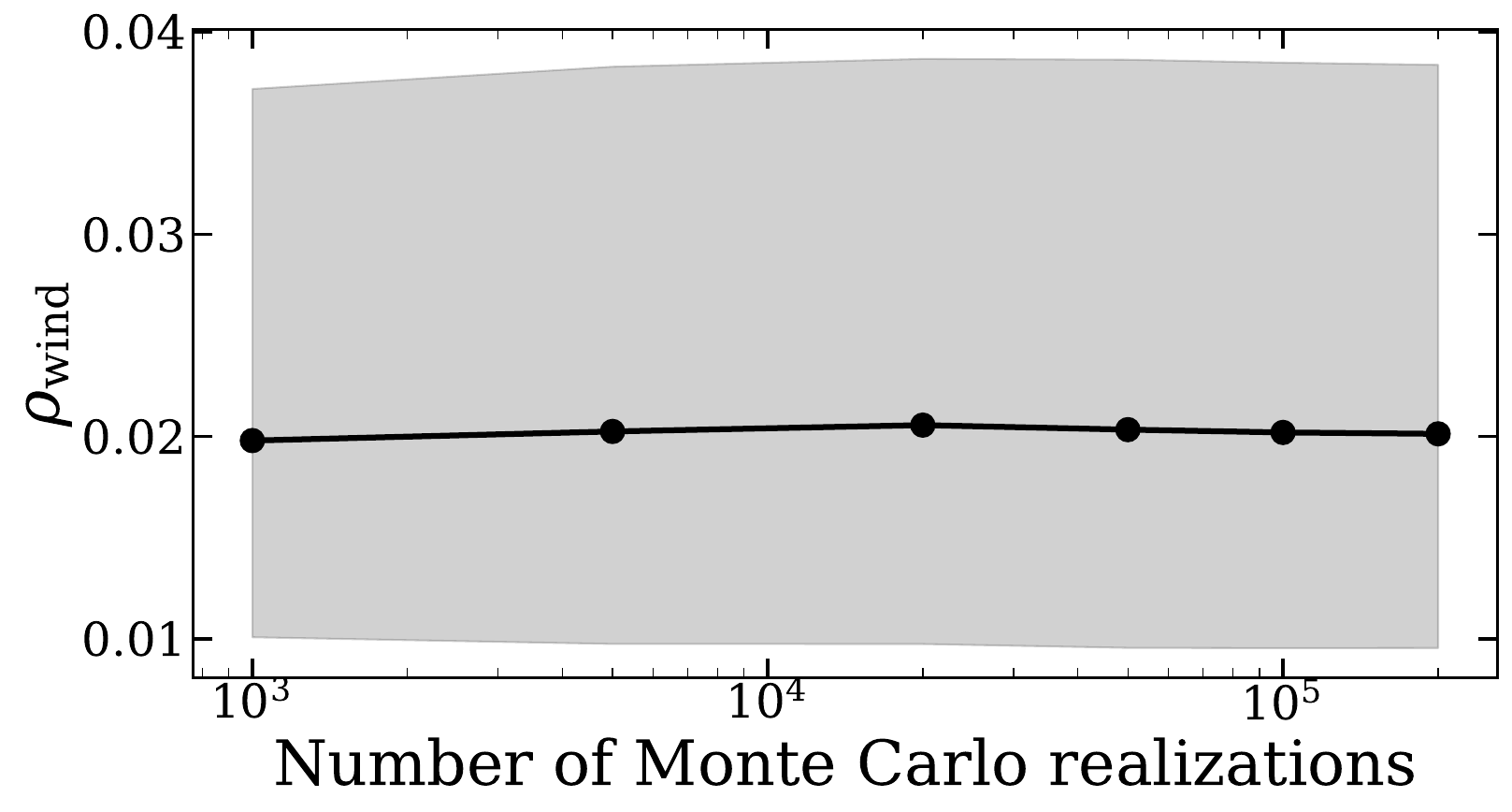}
\caption{Convergence of the median and 68 percent interval of $\rho_{\rm wind}$ with the number of Monte Carlo realizations.}
\label{fig:conv}
\end{figure}

All uncertainties are propagated with $2\times10^5$ realizations, using \texttt{numpy.random.default\_rng} with deterministic seeds; this realization count and seed policy apply to every posterior quoted in the paper. The priors are:
\begin{itemize}
\item $\xi_{\rm CR}$: log-uniform over $0.05$--$0.20$;
\item $f_{\pi^0}$: uniform over $0.30$--$0.37$;
\item $f_{\rm band}$: uniform over $0.65$--$0.85$;
\item measured efficiencies and luminosities: two-piece normal with the published asymmetric errors, truncated at zero by resampling rather than clipping;
\item starburst $\gamma$-ray luminosities: $0.10$~dex; starburst mechanical powers: $0.30$~dex;
\item shock radius and speed: log-uniform over the ranges quoted in Sect.~\ref{sec:fcal}.
\end{itemize}
Because $\eta_{\max}$ is common to the wind and starburst estimates, the ratio $\rho_{\rm starburst}/\rho_{\rm wind}$ is formed realization by realization, and its uncertainty cancels. The $0.34$~dex calibration scatter of the star-formation calorimeter enters only through the subtraction of Sect.~\ref{sec:two}; repeating that calculation with the scatter set to $0$, to $0.34$ and to $0.68$~dex gives 95 percent intervals of width $1.86$, $1.96$ and $2.18$~dex respectively, with medians $1.75$, $1.74$ and $1.72$. It is therefore a subdominant contribution to the tails. The median $\rho_{\rm wind}$ changes by $1.0$ percent over the last decade in the number of realizations, which confirms that the quoted percentiles are converged (Fig.~\ref{fig:conv}).

\subsection{Dispersion in \texorpdfstring{$\rho$}{rho} from the censored limits}

Let $\rho_i$ be log-normally distributed with dispersion $\sigma$ and a mean fixed by the stacking result, $\mathbb{E}[\rho]=\rho_{\rm wind}$, which implies $\mu=\ln\rho_{\rm wind}-\sigma^2/2$. Each source contributes a factor $\Phi[(\ln\rho_i^{95\%}-\mu)/\sigma]$ to the likelihood, where $\rho_i^{95\%}$ is the calorimetric ratio allowed by its own upper limit (Table~\ref{tab:swind}). The likelihood increases monotonically with $\sigma$ over most of the range, because a broad distribution lets a few sources carry the stacked signal while all eleven limits are satisfied; only a lower bound is therefore obtained, $\sigma>1.0$~dex at $\Delta(-2\ln\mathcal{L})=1$ and $\sigma>0.3$~dex at $\Delta(-2\ln\mathcal{L})=3.84$. Treating the 95 percent limits as hard constraints is conservative in the sense that it maximizes the tension and therefore the inferred dispersion; relaxing them would weaken the bound.

\section{Robustness of the comparisons}
\label{app:robust}

\subsection{Independence of the inter-class ratio from the acceleration parameters}

The absolute calorimetric ratios depend on $\eta_{\max}=\xi_{\rm CR}f_{\pi^0}f_{\rm band}$ and therefore on three quantities that are not measured here. The ratio between two classes does not, because the same $\eta_{\max}$ divides both:
\begin{equation}
\frac{\rho_{\rm starburst}}{\rho_{\rm wind}}
=\frac{\eta_{\rm starburst}/\eta_{\max}}{\eta_{\rm wind}/\eta_{\max}}
=\frac{\eta_{\rm starburst}}{\eta_{\rm wind}}.
\end{equation}
We verified this numerically: recomputing the whole chain with $\xi_{\rm CR}=0.30$ raises $\eta_{\max}$ to $3.1\times10^{-2}$ for the corresponding parameter draw and lowers $\rho_{\rm wind}$ to $0.017$, while the inter-class ratio remains $14.2$ to the last digit carried. All comparative statements in this paper are therefore insensitive to the acceleration physics; only the statements about absolute calorimetry are not, and they are always qualified by the assumed $\xi_{\rm CR}$.

\subsection{Sample comparability}

Two-sample Kolmogorov--Smirnov tests between the S-WIND and S-XRAY host samples give $D=0.26$, $p=0.52$ for $\log M_\bullet$; $D=0.55$, $p=0.007$ for distance; and $D=0.57$, $p=0.005$ for $\log L_{\rm bol}$ (adopting $\kappa_{14-195}=8$ for S-XRAY). The samples are matched in black-hole mass and not in distance or AGN power, which is why the comparison of Sect.~\ref{sec:two} is normalized by $L_{\rm Edd}$ and is presented as a null test.

\subsection{Leave-one-out tests}

Removing the two sources shared between the samples changes the mean $\log M_\bullet$ by $+0.13$~dex (S-WIND) and $-0.02$~dex (S-XRAY), and removes $TS=0.56$ from the S-XRAY total of $30.6$. Removing NGC~4151, the single object that dominates the S-WIND stacked significance, or NGC~7582, the object that drives the dispersion bound, changes no conclusion: the published stacking excluding NGC~4151 still gives $TS=15.1$ with a consistent flux and index \citep{Ajello2021}, and the dispersion bound of Sect.~\ref{sec:targets} is a lower limit that can only weaken.

\section{Class-specific parameters, stacking weights, and the censoring model}
\label{app:referee}

\subsection{If the two classes do not share acceleration parameters}
\label{app:xidiff}

The cancellation of $\xi_{\rm CR}$, $f_{\pi^0}$ and $f_{\rm band}$ in the inter-class ratio
(Appendix~\ref{app:robust}) holds only if the same values apply to both classes. Relaxing that,
we draw the three parameters independently for the wind and starburst populations over the
ranges of Appendix~\ref{app:mc}. The contrast becomes $\contrastindep$
($68\%$: $\contrastindeplo$--$\contrastindephi$), with
$P(\rho_{\rm starburst}>3\,\rho_{\rm wind})=\Pcontrastthree$: the median is unchanged, because the
priors are the same for both classes, but the interval widens.

The more useful statement is the inverse one. The measured ratio of apparent efficiencies is
$\rhoratio$; the contrast in $\rho$ is therefore erased entirely only if AGN-wind shocks convert
mechanical power into cosmic rays a factor $1/\xiratioone\simeq13$ less efficiently than
supernova shocks, and falls below three only for a factor $1/\xiratiothree\simeq4.5$. We are not
aware of a reason to expect a differential of that size, but the reader who does expect one can
read the contrast off directly, since it scales linearly with $\xi_{\rm wind}/\xi_{\rm starburst}$.
As for $f_{\rm band}$, the two classes have measured photon indices that agree within their
uncertainties ($\Gamma=2.1\pm0.3$ for S-WIND against $2.2$--$2.3$ for the starbursts), and the
band correction is thus common to within the spread those indices allow; it is nonetheless inherited
from the star-forming-galaxy calibration of \citet{Kornecki2020} and is not independently
measured for AGN winds.

\subsection{What the stacked efficiency constrains}
\label{app:weights}

The published efficiency is consistent with the ratio of the sample-mean luminosity to the
sample-mean kinetic power, $7.9\times10^{40}/1.8\times10^{44}$, and not with the arithmetic
mean of the individual ratios. A ratio-of-means estimator constrains the $\dot{E}$-weighted mean
$\mathbb{E}[\rho\dot{E}]/\mathbb{E}[\dot{E}]$, which equals $\mathbb{E}[\rho]$ if and only if
$\rho$ is uncorrelated with $\dot{E}$. The weights are concentrated: an effective sample size of
$\Neff$ of eleven, with $\topweight$ percent of the weight on PG~1211+143 alone.

The censored analysis of Appendix~\ref{app:mc} was therefore repeated imposing
$\rho\propto\dot{E}^{\,g}$ and normalizing the weighted mean to the published value. For
$g=0$ we recover $\sigma_{\log\rho}>\sigmarho$~dex; for $g=-0.5$ the bound tightens to
$\sigmaneg$~dex; for $g=+0.5$ it weakens to $\sigmapos$~dex, i.e.\ it disappears. The dispersion
bound therefore holds if $\rho$ is uncorrelated with, or anti-correlated with, the wind kinetic
power, and not if the two are strongly positively correlated. This is the weakest of the
constraints presented here, and Sect.~\ref{sec:targets} is worded accordingly.

\subsection{Hard versus soft treatment of the upper limits}
\label{app:censor}

Treating each published 95 percent limit as a hard threshold is an approximation forced by the
absence of published per-source likelihood profiles. Its direction can be tested with a soft likelihood in which every non-detection is a Gaussian measurement of zero
flux with $\sigma=F^{95\%}/1.645$, marginalised over the log-normal population. This gives
$\sigma_{\log\rho}>\sigmasoft$~dex, a stronger bound than the hard-threshold
$\sigmarho$~dex; the hard treatment is thus the conservative one. In addition,
of mock populations drawn with a true dispersion of $0.2$~dex only $\compatlow$ percent satisfy
all eleven published limits simultaneously, against $\compathigh$ percent at $1.0$~dex.

\section{Sensitivity to the spectral extrapolation}
\label{app:gamma}

The band corrections of Sect.~\ref{sec:homog} extrapolate outside the measured ranges. Varying $\Gamma$ over the published $1\sigma$ intervals ($2.1\pm0.3$ for S-WIND, $2.32^{+0.40}_{-0.35}$ for S-XRAY) changes $k$ by $-13/+16$ percent and $-18/+21$ percent respectively. The largest single effect on any conclusion is on the ratio of Sect.~\ref{sec:two}, which moves from $1.75$ to between $1.4$ and $2.2$; it remains consistent with unity in every case.

\end{document}